\documentclass[draft]{agujournal2018}
\draftfalse
\usepackage{url}
\usepackage{lineno}
\usepackage{booktabs}
\usepackage{subcaption}
\usepackage{journal_macros}
\usepackage{changepage}
\usepackage{amssymb}
\usepackage{multirow}

\usepackage{enumitem}
\setlist{noitemsep,topsep=1pt}

\usepackage[bookmarks=false, pdfnewwindow=true, colorlinks=true, linkcolor=magenta, citecolor=blue, filecolor=cyan, urlcolor=purple, final=true]{hyperref}

\journalname{Space Weather}

\graphicspath{{./}{figures/}}

\begin{document}

\title{CMEs and SEPs During November--December 2020: A Challenge for Real-Time Space Weather Forecasting}

\authors{Erika~Palmerio\affil{1,2}\thanks{Now at Predictive Science Inc., San Diego, CA 92121, USA}, Christina~O.~Lee\affil{1}, M.~Leila~Mays\affil{3}, Janet~G.~Luhmann\affil{1}, David~Lario\affil{3}, Beatriz~S{\'a}nchez-Cano\affil{4}, Ian~G.~Richardson\affil{5,3}, Rami~Vainio\affil{6}, Michael~L.~Stevens\affil{7}, Christina~M.~S.~Cohen\affil{8}, Konrad~Steinvall\affil{9,10}, Christian~M{\"o}stl\affil{11}, Andreas~J.~Weiss\affil{11,12}, Teresa~Nieves-Chinchilla\affil{3}, Yan~Li\affil{1}, Davin~E.~Larson\affil{1}, Daniel~Heyner\affil{13}, Stuart~D.~Bale\affil{1,14}, Antoinette~B.~Galvin\affil{15}, Mats~Holmstr{\"o}m\affil{16}, Yuri~V.~Khotyaintsev\affil{9}, Milan~Maksimovic\affil{17}, and Igor~G.~Mitrofanov\affil{18}}

\begin{adjustwidth}{-1cm}{-1cm}
\affiliation{1}{Space Sciences Laboratory, University of California--Berkeley, Berkeley, CA 94720, USA}
\affiliation{2}{CPAESS, University Corporation for Atmospheric Research, Boulder, CO 80301, USA}
\affiliation{3}{Heliophysics Science Division, NASA Goddard Space Flight Center, Greenbelt, MD 20771, USA}
\affiliation{4}{School of Physics and Astronomy, University of Leicester, Leicester, LE1 7RH, UK}
\affiliation{5}{Department of Astronomy, University of Maryland, College Park, MD 20742, USA}
\affiliation{6}{Department of Physics and Astronomy, University of Turku, FI-20014 Turku, Finland}
\affiliation{7}{Harvard--Smithsonian Center for Astrophysics, Cambridge, MA 02138, USA}
\affiliation{8}{California Institute of Technology, Pasadena, CA 91125, USA}
\affiliation{9}{Swedish Institute of Space Physics, S-75121 Uppsala, Sweden}
\affiliation{10}{Department of Physics and Astronomy, Uppsala University, S-75120 Uppsala, Sweden}
\affiliation{11}{Space Research Institute, Austrian Academy of Sciences, A-8042 Graz, Austria}
\affiliation{12}{Institute of Physics, University of Graz, A-8010 Graz, Austria}
\affiliation{13}{Institut f{\"u}r Geophysik und extraterrestrische Physik, TU Braunschweig, D-38106 Braunschweig, Germany}
\affiliation{14}{Physics Department, University of California--Berkeley, Berkeley, CA 94720, USA}
\affiliation{15}{Space Science Center, University of New Hampshire, Durham, NH 03824, USA}
\affiliation{16}{Swedish Institute of Space Physics, S-98192 Kiruna, Sweden}
\affiliation{17}{LESIA, Observatoire de Paris, Univ. PSL, CNRS, Sorbonne Univ., Univ. Paris Diderot, F-92195 Meudon, France}
\affiliation{18}{Institute for Space Research, RU-117997 Moscow, Russian Federation}
\end{adjustwidth}

\correspondingauthor{E. Palmerio}{epalmerio@predsci.com}


\begin{keypoints}
\item We model the inner heliospheric context between the two eruptive flares of 2020 Nov 29 and Dec 7 using the WSA--Enlil--SEPMOD modelling chain
\item CME input parameters are obtained using remote-sensing observations from Earth and STEREO-A, located near the L5 point
\item The modelling setup used in this work can provide useful CME and shock-accelerated SEP predictions in real-time forecasts
\end{keypoints}


\begin{abstract}
Predictions of coronal mass ejections (CMEs) and solar energetic particles (SEPs) are a central issue in space weather forecasting. In recent years, interest in space weather predictions has expanded to include impacts at other planets beyond Earth as well as spacecraft scattered throughout the heliosphere. In this sense, the scope of space weather science now encompasses the whole heliospheric system, and multi-point measurements of solar transients can provide useful insights and validations for prediction models. In this work, we aim to analyse the whole inner heliospheric context between two eruptive flares that took place in late 2020, i.e.\ the M4.4 flare of November~29 and the C7.4 flare of December~7. This period is especially interesting because the STEREO-A spacecraft was located ${\sim}60^{\circ}$ east of the Sun--Earth line, giving us the opportunity to test the capabilities of ``predictions at $360^{\circ}$'' using remote-sensing observations from the Lagrange L1 and L5 points as input. We simulate the CMEs that were ejected during our period of interest and the SEPs accelerated by their shocks using the WSA--Enlil--SEPMOD modelling chain and four sets of input parameters, forming a ``mini-ensemble''. We validate our results using in-situ observations at six locations, including Earth and Mars. We find that, despite some limitations arising from the models' architecture and assumptions, CMEs and shock-accelerated SEPs can be reasonably studied and forecast in real time at least out to several tens of degrees away from the eruption site using the prediction tools employed here.
\end{abstract}


\begin{plainsummary}
Coronal mass ejections (CMEs) and solar energetic particles (SEPs) are phenomena from the Sun that are able to cause significant disturbances at Earth and other planets. Reliable predictions of these events and their effects are amongst the major goals of space weather forecasts, which aim to tackle all processes related to solar activity that can endanger human technology and society. In recent years, the breadth of space weather science has started to encompass other locations than Earth, ranging from all solar system planets to spacecraft scattered throughout space. In this work, we test our current capabilities in predicting space weather events in the inner solar system (i.e., within the orbit of Mars) for a period in late 2020. We use a chain of models that are able to simulate the background solar wind as well as transient phenomena such as CMEs and SEPs, and compare our results with spacecraft measurements from six well-separated locations, including Earth and Mars. We find that our current forecasting tools, despite their limitations, can successfully provide reasonable predictions of both CMEs and SEPs, especially out to several tens of degrees around the corresponding eruption source region on the Sun.
\end{plainsummary}


\section{Introduction} \label{sec:intro}

Coronal mass ejections \citep[CMEs; e.g.,][]{webb2012} and solar energetic particles \citep[SEPs; e.g.,][]{reames2021} are manifestations of the variable solar activity and important drivers of space weather effects at Earth and other planets \citep[e.g.,][]{eastwood2017,koskinen2017,temmer2021}. A CME impacting Earth's magnetosphere may trigger a geomagnetic storm depending on its kinematic and magnetic properties \citep[e.g.,][]{gonzalez1994,zhang2007}. SEPs accelerated by solar flares and/or CME-driven shocks are able to reach energies of tens of MeV/nuc or higher and may penetrate Earth's ionosphere \citep[e.g.,][]{klein2017,malandraki2018} or even produce secondary particles that reach the ground \citep[causing a so-called `ground level enhancement' or GLE; e.g.,][]{nitta2012,shea2012}. These phenomena frequently occur in concert, especially in the case of large and fast CMEs that are associated with strong flares and that are able to accelerate particles at the shocks ahead of them, generating so-called `gradual' SEP events \citep[e.g.,][]{desai2016}. Amongst the active areas of CME and SEP research, significant efforts are dedicated to improving forecasts of their arrival and space weather impact at a given location in the heliosphere. Recent reviews on the current status of space weather science were compiled by \citet{vourlidas2019} for CMEs and by \citet{anastasiadis2019} for SEPs.

The interplanetary propagation of CMEs has been addressed during the past couple of decades by the development of a multitude of models, ranging from numerical \citep[based on magnetohydrodynamics or MHD equations, e.g.,][]{odstrcil1999,pomoell2018,wu2007} to analytical \citep[e.g.,][]{coronaromero2017,hess2015,vrsnak2013} to empirical \citep[e.g.,][]{gopalswamy2001,paouris2017,vrsnak2007}. The plethora of existing prediction tools has drawn the space weather community towards coordinated, international efforts (see, e.g., the CME Arrival Time and Impact Working Team: \url{https://ccmc.gsfc.nasa.gov/iswat/CME-Arrival-Time-Working-Team}) focused on benchmarking CME forecasts, especially in terms of hit/miss and arrival time \citep[e.g.,][]{riley2018,verbeke2019a,wold2018,zhao2004}. These studies found out that, despite continued efforts, typical uncertainties for arrival time are still of the order of ${\pm}10$~hours and typical accuracies in determining a hit or a miss lie around 85\%. Predictions of the magnetic structure of CMEs, also known as $B_{Z}$ forecasts, are currently even less advanced \citep[e.g.,][]{kilpua2019a}, although this issue has gained significant momentum during recent years and is being tackled with MHD \citep[e.g.,][]{jin2017,shiota2016,verbeke2019b}, analytical \citep[e.g.,][]{isavnin2016, kay2022, mostl2018}, and machine learning \citep[e.g.,][]{reiss2021} models.

Predictions of SEPs have also been explored with a number of physics-based \citep[e.g.,][]{marsh2015,schwadron2010,zhang2017}, empirical \citep[e.g.,][]{anastasiadis2017,bruno2021,posner2007,richardson2018b}, and machine learning \citep[e.g.,][]{kasapis2022,lavasa2021,stumpo2021} models, some of which can additionally be coupled with coronal and/or heliospheric MHD simulations to investigate time-dependent particle acceleration \citep[e.g.,][]{luhmann2017b,wijsen2021,young2021}. International efforts (see, e.g., the SEP Validation Team: \url{https://ccmc.gsfc.nasa.gov/iswat/SEP-Validation-Team}) are underway to assess the current status of SEP forecasting and to establish community-wide metrics. Recently, \citet{bain2021} assessed the performance of the National Oceanic and Atmospheric Administration's Space Weather Prediction Center (NOAA/SWPC) proton event forecasts, finding for ${\geq}10$~MeV (${\geq}100$~MeV) proton warnings a probability of detection of 91\% (53\%) and a false alarm ratio of 24\% (38\%), with a median lead time of 88~min (10~min). The authors suggested that these results may serve as a benchmark for SEP models that can operate in a nowcast setting, and noted that a particular challenge is to accurately predict the onset, peak, and end times and fluxes of SEP events in different energy ranges.

In addition to dedicated efforts in the context of space weather, which are traditionally focused on the Sun--Earth chain, the research community has also taken advantage of data from multiple spacecraft forming part of the Heliophysics System Observatory to perform multi-point studies of CMEs \citep[e.g.,][]{burlaga1981,davies2020,palmerio2021d,witasse2017} and SEPs \citep[e.g.,][]{bruno2021,lee2018,richardson2014,rodriguezgarcia2021} at well-separated locations in both longitude and heliocentric distance. Such works have uncovered myriad characteristics and processes that may not be identified based on single-viewpoint in-situ measurements or require statistical surveys of large samples of events. These include the detection of longitudinal variations in the shape of CME-driven shock fronts and/or the magnetic structure of flux ropes within CMEs  \citep[e.g.,][]{farrugia2011,kilpua2011,mulligan2013}, the characterisation of the radial evolution of CMEs \citep[e.g.,][]{good2019,salman2020,vrsnak2019}, the observation of unexpectedly large longitudinal spreads for some SEP events \citep[e.g.,][]{dresing2012,gomezherrero2015,palmerio2021a,wiedenbeck2013}, and the investigation of the longitudinal variation of SEP profiles and properties \citep[e.g.,][]{cohen2014,lario2006,lario2013,lario2016,rouillard2012}. Furthermore, multi-point data for a single event can provide useful constraints and validation for CME and SEP propagation models.

An additional level of complexity is given by periods that are characterised by several, successive eruptions \citep[see][for a review]{lugaz2017a}, which can lead to solar wind as well as interplanetary magnetic field preconditioning and thus result in altered (sometimes enhanced) CME geoeffectiveness \citep[e.g.,][]{liu2014,scolini2020} and higher SEP fluxes \citep[e.g.,][]{lario2014,richardson1996}. In this regard, studies that aim to analyse and/or model successive CME and SEP events measured at different, well-separated locations in the inner heliosphere \citep[e.g.,][]{bain2016,dumbovic2019,luhmann2018,mostl2012,palmerio2019} undertake the holistic approach necessary to better understand the whole heliospheric context and the interplay of complex processes. The recent years have seen the launch of three missions (i.e., Parker Solar Probe, BepiColombo, and Solar Orbiter) that are currently cruising and sampling the solar wind in the inner heliosphere. These ``newer'' spacecraft, together with the ``older'' but still operational ones (i.e., STEREO-A and those near Earth and Mars), represent an excellent opportunity to test, validate, and improve our current modelling capabilities and understanding of the inner heliosphere ``at $360^{\circ}$''.

\begin{figure}[ht]
\centering
\includegraphics[width=.9\linewidth]{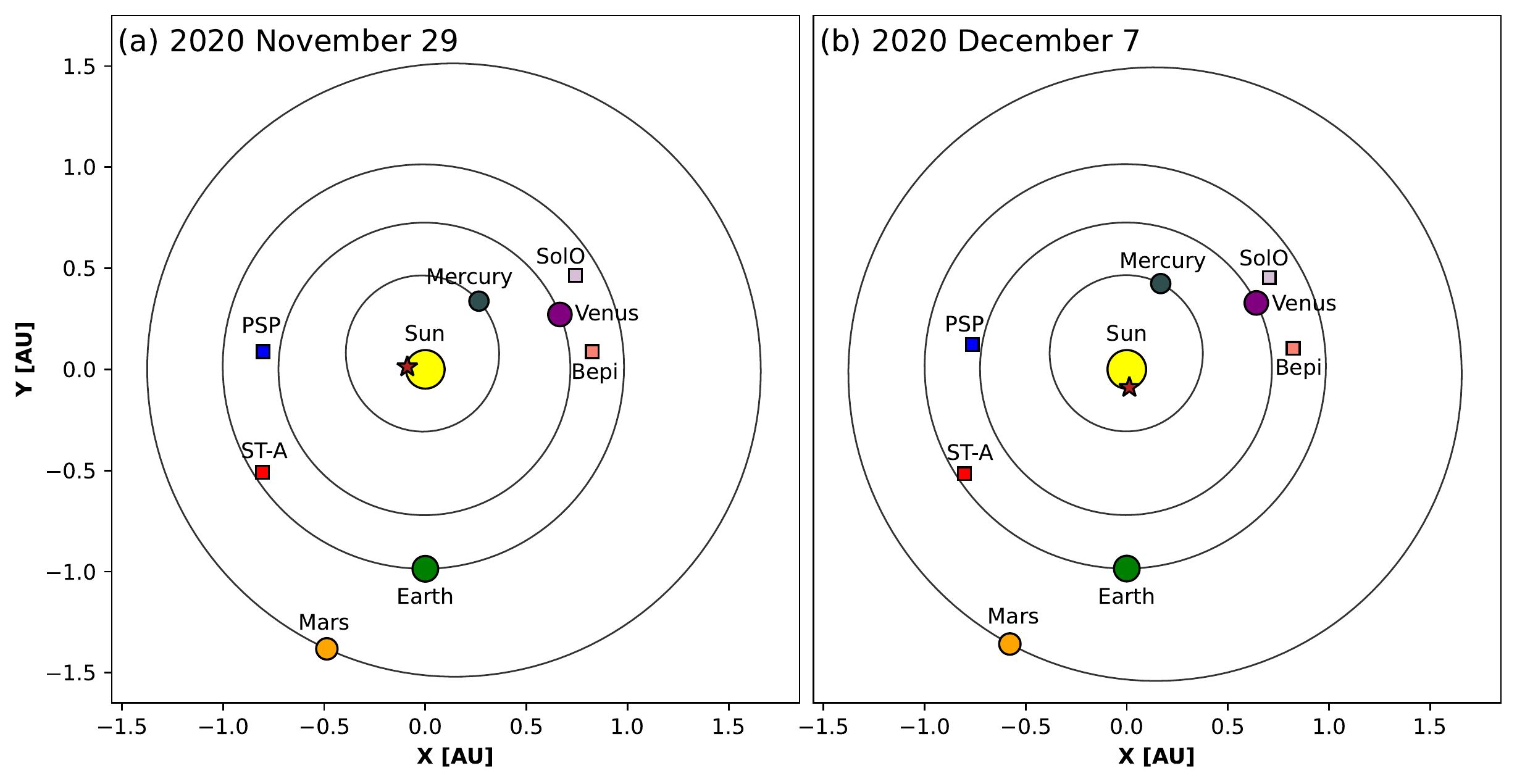}
\caption{Position of planets and spacecraft in the inner heliosphere on (a) 2020 November 29 and (b) 2020 December 7. The orbits of all planets are also indicated. The locations of (a) Flare1 and (b) Flare2 are marked with star symbols on the surface of the Sun. Bepi = BepiColombo; PSP = Parker Solar Probe; SolO = Solar Orbiter; ST-A = STEREO-A.}
\label{fig:orbits}
\end{figure}

Using this framework, the aim of this article is to use observations and modelling to understand the heliospheric context (including CMEs and SEPs) following two noteworthy eruptive flares that took place in late 2020, namely the M4.4 flare of November~29 (originating from S25E96 in Stonyhurst coordinates and peaking at 13:11~UT) and the C7.4 flare of December~7 (originating from S25W08 in Stonyhurst coordinates and peaking at 16:32~UT)---hereafter, Flare1 and Flare2. Both commenced from the same source region, AR~12790, which was still behind the Earth-facing eastern limb at the time of Flare1, indicating that its ``true'' X-ray class might have been even higher than observed. The first of these eruptions has already gained significant attention, since it was associated with the first widespread SEP event of Solar Cycle 25 and with a CME detected in situ by the Parker Solar Probe and STEREO-A spacecraft \citep[e.g.,][]{cohen2021, giacalone2021, kollhoff2021, kouloumvakos2022, lario2021, mason2021, mitchell2021, mostl2022, nieveschinchilla2022}. The configurations of planets and spacecraft within the orbit of Mars on the days of the two flares are shown in Figure~\ref{fig:orbits}. An interesting aspect is that STEREO-A was located ${\sim}60^{\circ}$ east of Earth, thus providing the opportunity to test modelling and forecasting capabilities based on observations from the Lagrange L1 (Earth) and L5 (STEREO-A) points. This article is organised as follows: In Section~\ref{sec:data}, we provide an overview of the spacecraft instrumentation and data sets employed in this study. In Section~\ref{sec:setup}, we describe the the simulation setup and the different sets of CME input parameters used to generate a ``mini-ensemble'' of model runs of the inner heliosphere. In Section~\ref{sec:results}, we present the simulation results and compare them with in-situ solar wind and SEP observations from six locations. In Section~\ref{sec:discussion} we discuss the implications of our results for space weather forecasting and in Section~\ref{sec:conclusions} we summarise our study and draw our conclusions.

\section{Spacecraft Data} \label{sec:data}

The observations used in this study include remote-sensing observations from two viewpoints (Earth and STEREO-A, representing views from L1 and L5, respectively) and in-situ measurements from six locations (Parker Solar Probe, STEREO-A, Mars, Earth, BepiColombo, and Solar Orbiter) as shown in Figure~\ref{fig:orbits}. The corresponding heliocentric distances throughout the analysed period are approximately: 0.79~AU for Parker Solar Probe, 0.96~AU for STEREO-A, 1.47~AU for Mars, 0.99~AU for Earth, 0.83~AU for BepiColombo, and 0.86~AU for Solar Orbiter. Remote-sensing observations include extreme ultra-violet (EUV) solar disc and white-light (WL) coronagraph data, whilst in-situ measurements include, where available, magnetic field, plasma, and energetic particle data. In the following, we refer to SEPs of energy in the range ${\sim}1$--$10$~MeV as ``lower-energy particles'' and those of energy ${\gtrsim}10$~MeV as ``higher-energy particles''. All particle data employed here refer to protons.

\begin{itemize}

\item \textbf{Remote-sensing observations:} EUV images of the solar disc wavelengths are provided by the Atmospheric Imaging Assembly \citep[AIA;][]{lemen2012} onboard the Solar Dynamics Observatory \citep[SDO;][]{pesnell2012} orbiting Earth and the Extreme UltraViolet Imager (EUVI) part of the Sun Earth Connection Coronal and Heliospheric Investigation \citep[SECCHI;][]{howard2008a} suite onboard the Solar Terrestrial Relations Observatory \citep[STEREO;][]{kaiser2008} Ahead (STEREO-A) spacecraft. WL coronagraph imagery is supplied by the C2 and C3 telescopes of the Large Angle and Spectrometric Coronagraph \citep[LASCO;][]{brueckner1995} onboard the Solar and Heliospheric Observatory \citep[SOHO;][]{domingo1995} at Earth's Lagrange L1 point and the COR2 camera of SECCHI onboard STEREO-A.

\item \textbf{Parker Solar Probe:} At the Parker Solar Probe \citep[PSP;][]{fox2016} spacecraft, we use magnetic field measurements from the fluxgate magnetometer part of the FIELDS \citep{bale2016} investigation and plasma data from the Solar Probe Cup \citep[SPC;][]{case2020} part of the Solar Wind Electrons Alphas and Protons \citep[SWEAP;][]{kasper2016} investigation. SEP measurements come from the Energetic Particle Instrument-High \citep[EPI-Hi;][]{wiedenbeck2017} part of the Integrated Science Investigation of the Sun \citep[IS$\odot$IS;][]{mccomas2016}: lower-energy particles are detected by the Low Energy Telescope A (LETA), whereas higher-energy particles are detected by the High Energy Telescope A (HETA).

\item \textbf{STEREO-A:} At the STEREO-A spacecraft, we use magnetic field measurements form the Magnetic Field Experiment \citep[MFE;][]{acuna2008}, lower-energy particle data from the Low Energy Telescope \citep[LET;][]{mewaldt2008}, and higher-energy particle data from the High Energy Telescope \citep[HET;][]{vonrosenvinge2008}, all part of the In situ Measurements of Particles And CME Transients \citep[IMPACT;][]{luhmann2008} investigation. In-situ plasma data come from the Plasma and Suprathermal Ion Composition \citep[PLASTIC;][]{galvin2008} instrument.

\item \textbf{Mars:} In-situ measurements from Mars orbit are provided by three spacecraft. Plasma data come from the Analyzer of Space Plasmas and Energetic Atoms \citep[ASPERA-3;][]{barabash2006} onboard Mars Express \citep[MEX;][]{chicarro2004}. Lower-energy particle data are supplied by the Solar Energetic Particle \citep[SEP;][]{larson2015} investigation onboard the Mars Atmosphere and Volatile Evolution \citep[MAVEN;][]{jakosky2015} spacecraft. Higher-energy particles are not directly measured by the spacecraft orbiting Mars; nevertheless, we use proxy data from the High Energy Neutron Detector (HEND) part of the Gamma Ray Spectrometer \citep[GRS;][]{boynton2004} suite onboard 2001 Mars Odyssey \citep[MOdy;][]{saunders2004} as well as MAVEN/SEP. Magnetic field measurements of the upstream solar wind from MAVEN are not available during the period under study.

\item \textbf{Earth:} In-situ measurements from near Earth are provided by three spacecraft located at Earth's Lagrange L1 point. Magnetic field and plasma data are supplied by the Magnetic Field Investigation \citep[MFI;][]{lepping1995} and Solar Wind Experiment \citep[SWE;][]{ogilvie1995} instruments onboard Wind \citep{ogilvie1997}. Lower-energy particles are detected by the Electron, Proton, and Alpha Monitor \citep[EPAM;][]{gold1998} onboard the Advanced Composition Explorer \citep[ACE;][]{stone1998}. Higher-energy particles are measured by the Energetic and Relativistic Nuclei and Electron \citep[ERNE;][]{torsti1995} experiment onboard SOHO.

\item \textbf{BepiColombo:}  The BepiColombo \citep[Bepi;][]{benkhoff2010,milillo2020} spacecraft is currently in cruise phase before its planned Mercury orbit insertion in 2025. Magnetic field measurements are provided by the Mercury Planetary Orbiter (MPO) Magnetometer \citep[MPO-MAG;][]{heyner2021}. Data from the particle spectrometer are not available during the period under study; nevertheless, we use in our investigation proton counts over both lower and higher SEP energy ranges from the Bepicolombo Environment Radiation Monitor \citep[BERM;][]{pinto2021}. The plasma instrument from BepiColombo is not operational during cruise phase.

\item \textbf{Solar Orbiter:} At the Solar Orbiter \citep[SolO;][]{muller2020} spacecraft, we use magnetic field data from the Magnetometer \citep[MAG;][]{horbury2020} and electron density measurements from the Radio and Plasma Waves \citep[RPW;][]{maksimovic2020} instrument (data from the instrument dedicated to solar wind plasma are not available during the events investigated here). We also include solar wind speed estimates based on MAG and RPW data using deHoffmann--Teller analysis, the details of which are described in \citet{steinvall2021}. SEPs are measured with the Energetic Particle Detector \citep[EPD;][]{rodriguezpacheco2020}, specifically with the Electron Proton Telescope (EPT) for lower-energy particles and the High Energy Telescope (HET) for higher-energy ones.

\end{itemize}


\section{Simulation Setup} \label{sec:setup}

To model the heliospheric conditions during the period under analysis, we use the Wang--Sheeley--Arge \citep[WSA;][]{arge2004} coronal model coupled with the 3D MHD Enlil \citep{odstrcil2003,odstrcil2004} heliospheric model, henceforth WSA--Enlil. As input for WSA, we employ time-dependent National Solar Observatory (NSO) Global Oscillation Network Group \citep[GONG;][]{harvey1996} zero-point-corrected magnetogram synoptic maps, used to generate solar wind and interplanetary magnetic field conditions up to the Enlil inner boundary, set at $21.5\,R_{\odot}$ or 0.1~AU. From there, Enlil uses WSA output as input to model the heliospheric conditions by solving the MHD equations using a flux-corrected-transport algorithm. Since we are interested in modelling CME propagation and SEP transport up to the orbit of Mars, we set the Enlil outer boundary to 3~AU. We then send the WSA--Enlil output data files to the SEPMOD \citep{luhmann2007,luhmann2010} SEP prediction code. SEPMOD uses field line and shock description outputs from WSA--Enlil to forward model SEP flux time series, typically for protons at energies from 1 to 100~MeV, along sequential observer-connected field lines using shock-source injections followed by a constant-energy, guiding-centre transport approximation. In our WSA--Enlil--SEPMOD simulation runs, we use WSA version 5.2, Enlil version 2.8f, and SEPMOD version 2. The runs are performed at NASA's Community Coordinated Modeling Center (CCMC).

CMEs are inserted in the model at the Enlil inner boundary as hydrodynamic, high-pressure pulses that are able to drive shock waves if their speed is sufficiently greater than that of the ambient solar wind. Because of the inherent limits of the numerical simulation approach used in Enlil, this inner boundary is set at $21.5\,R_{\odot}$. As a result, extrapolations of magnetograph and coronagraph data are needed to drive the model on that spherical ``source surface''.  For this reason, there are no coronal or solar wind disturbances---and, thus, no modeled interplanetary shocks---below $21.5\,R_{\odot}$. Note that since CMEs are inserted into the Enlil heliospheric domain as hydrodynamic structures, they lack internal magnetic fields. Hence, it is not possible to model their magnetic configuration, e.g., to predict the presence or absence of a flux rope structure in situ.

\begin{figure}[t!]
\centering
\hspace*{-0.1\linewidth}
\includegraphics[width=1.2\linewidth]{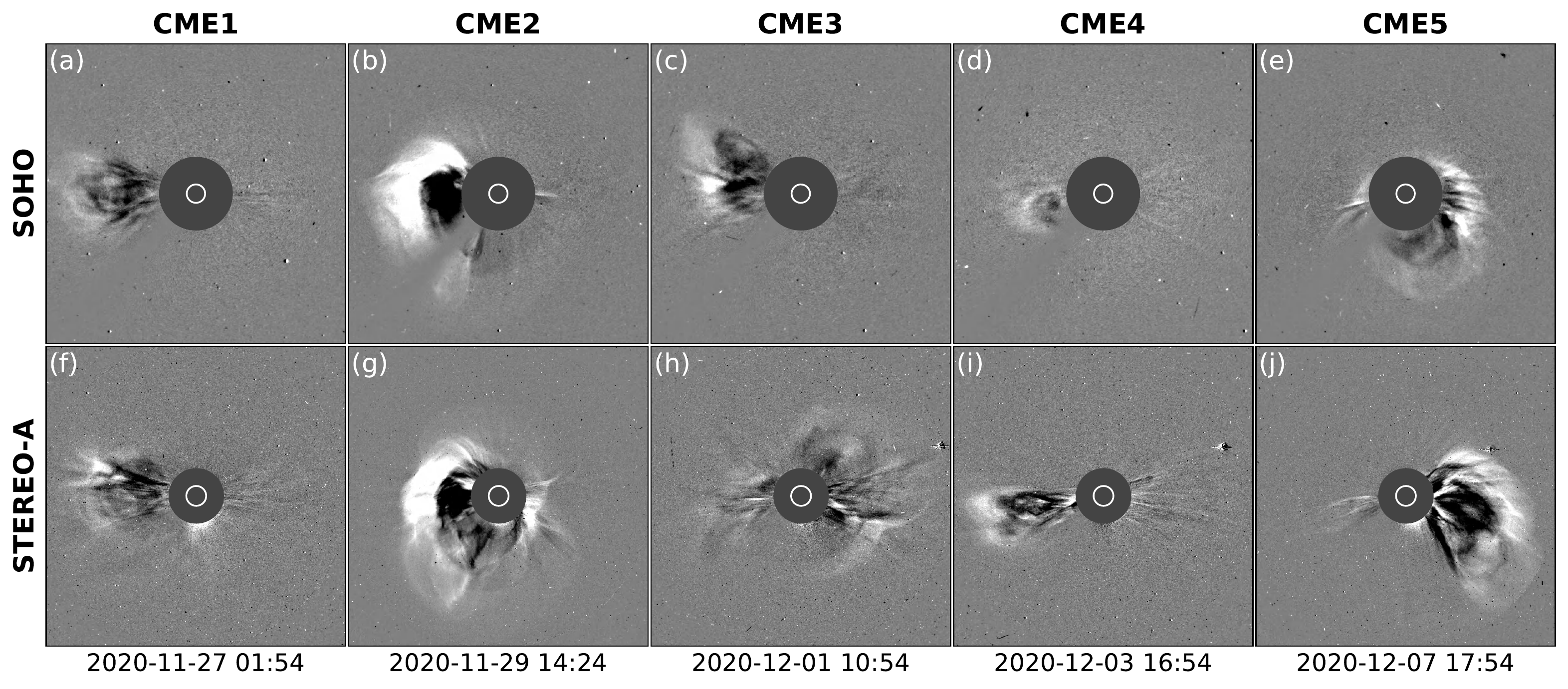}
\caption{The five CMEs modelled in this study as seen in coronagraph imagery from (top row) SOHO/LASCO/C3 at L1 and (bottom row) STEREO/SECCHI/COR2-A close to L5. Flare1 and Flare2 are associated with CME2 and CME5, respectively.}
\label{fig:coronagraph}
\end{figure}

In order to model the heliospheric context and solar wind transients during the time interval between Flare1 and Flare2, we selected all the CMEs \citep[excluding minor outflows and 'jet-like' eruptions, according to the definition of][]{vourlidas2013,vourlidas2017} observed between 2020 November~29 and December~7 via inspection of SOHO/LASCO and STEREO/SECCHI data. In addition, Flare1 was preceded by a similarly-directed eruption that originated late on 2020 November~26 \citep[e.g.,][]{lario2021, nieveschinchilla2022}. We also include this event in our investigation, since it likely led to preconditioning of the heliosphere, which may have affected the propagation speed of the CME associated with Flare1 and the transport of associated SEPs. In all, we selected five CMEs, shown in Figure~\ref{fig:coronagraph} from the perspectives of both SOHO and STEREO-A. CME2 and CME5 are associated with Flare1 and Flare2, respectively.  

To explore how variations to CME input parameters affect shock arrival times, as well as shock--observer connectivity and thus SEP access, we employ a ``mini-ensemble'' approach. The advantages of ensemble modelling techniques, where the outputs of a series of model runs with slightly different input conditions are compared, for space weather forecasting were briefly reviewed by \citet{knipp2016} and \citet{murray2018}. Multiple studies have performed  
ensemble modelling of CME propagation using WSA--Enlil \citep[e.g.,][]{cash2015,emmons2013,lee2013,lee2015,mays2015a,mays2015b,pizzo2015} as well as other models \citep[e.g.,][]{amerstorfer2018,amerstorfer2021,dumbovic2018,kay2020}. In this work, we explore how outcomes such as CME arrival time, interplanetary magnetic field connectivity, and SEPs spatial spread and peak fluxes vary with respect to CME input parameters. Our WSA--Enlil--SEPMOD mini-ensemble consists of four members, with the CME input parameters for each shown in Table~\ref{tab:cme_list} and derived in the following way:

\begin{table}[!p]
\centering
\caption{CME Input Parameters for the WSA--Enlil--SEPMOD Simulation Runs.}
\label{tab:cme_list}

\begin{subtable}{.99\linewidth}
\centering
\caption{Run1---Real time}
\begin{tabular}{c c c c c c c c}
\toprule
CME & $t_{0}$ & $\theta$  & $\phi$ & $\gamma$ & $\psi_{1}$ & $\psi_{2}$ & $V$ \\
\# & UT & [$^{\circ}$]  & [$^{\circ}$] & [$^{\circ}$] & [$^{\circ}$] & [$^{\circ}$] & [km$\cdot$s$^{-1}$]\\
\midrule
1 & 2020-11-27T03:43  & $6.0$ & $-100.0$ & $0.0$ & $32.00$ & $32.00$ & $550.0$\\
2 & 2020-11-29T16:00 & $-6.0$ & $-75.0$ & $0.0$ & $58.00$ & $58.00$ & $1336.0$\\
3 & 2020-12-01T10:59 & $8.0$ & $-45.0$ & $0.0$ & $35.00$ & $35.00$ & $774.0$\\
4 & 2020-12-03T22:40 & $-5.0$ & $-130.0$ & $0.0$ & $20.00$ & $20.00$ & $314.0$\\
5 & 2020-12-07T18:49 & $-21.0$ & $12.0$ & $0.0$ & $41.00$ & $41.00$ & $1383.0$\\
\bottomrule
\end{tabular}
\end{subtable}

\begin{subtable}{.99\linewidth}
\centering
\caption{Run2---Cone model}
\begin{tabular}{c c c c c c c c}
\toprule
CME & $t_{0}$ & $\theta$  & $\phi$ & $\gamma$ & $\psi_{1}$ & $\psi_{2}$ & $V$ \\
\# & UT & [$^{\circ}$]  & [$^{\circ}$] & [$^{\circ}$] & [$^{\circ}$] & [$^{\circ}$] & [km$\cdot$s$^{-1}$]\\
\midrule
1 & 2020-11-27T04:59  & $4.0$ & $-98.0$ & $0.0$ & $22.95$ & $22.95$ & $425.1$\\
2 & 2020-11-29T15:25 & $-4.0$ & $-88.0$ & $0.0$ & $60.46$ & $60.46$ & $1661.9$\\
3 & 2020-12-01T12:51 & $18.0$ & $-40.0$ & $0.0$ & $36.87$ & $36.87$ & $811.6$\\
4 & 2020-12-03T23:05 & $-6.0$ & $-135.0$ & $0.0$ & $17.46$ & $17.46$ & $328.5$\\
5 & 2020-12-07T19:38 & $-22.0$ & $5.0$ & $0.0$ & $42.07$ & $42.07$ & $1120.9$\\
\bottomrule
\end{tabular}
\end{subtable}

\begin{subtable}{.99\linewidth}
\centering
\caption{Run3---Croissant model}
\begin{tabular}{c c c c c c c c}
\toprule
CME & $t_{0}$ & $\theta$  & $\phi$ & $\gamma$ & $\psi_{1}$ & $\psi_{2}$ & $V$ \\
\# & UT & [$^{\circ}$]  & [$^{\circ}$] & [$^{\circ}$] & [$^{\circ}$] & [$^{\circ}$] & [km$\cdot$s$^{-1}$]\\
\midrule
1 & 2020-11-27T04:30  & $4.0$ & $-96.0$ & $-25.0$ & $39.67$ & $20.49$ & $521.8$\\
2 & 2020-11-29T15:27 & $-8.0$ & $-92.0$ & $-38.0$ & $58.75$ & $35.45$ & $1700.6$\\
3 & 2020-12-01T13:33 & $15.0$ & $-40.0$ & $-20.0$ & $50.33$ & $29.99$ & $599.1$\\
4 & 2020-12-03T19:52 & $-9.0$ & $-155.0$ & $5.0$ & $34.63$ & $15.07$ & $502.5$\\
5 & 2020-12-07T19:36 & $-22.0$ & $8.0$ & $20.0$ & $51.53$ & $31.33$ & $1120.9$\\
\bottomrule
\end{tabular}
\end{subtable}

\begin{subtable}{.99\linewidth}
\centering
\caption{Run4---A posteriori}
\begin{tabular}{c c c c c c c c}
\toprule
CME & $t_{0}$ & $\theta$  & $\phi$ & $\gamma$ & $\psi_{1}$ & $\psi_{2}$ & $V$ \\
\# & UT & [$^{\circ}$]  & [$^{\circ}$] & [$^{\circ}$] & [$^{\circ}$] & [$^{\circ}$] & [km$\cdot$s$^{-1}$]\\
\midrule
1 & 2020-11-27T04:59  & $4.0$ & $-98.0$ & $0.0$ & $23.00$ & $23.00$ & $400.0$\\
2 & 2020-11-29T15:25 & $-4.0$ & $-96.0$ & $0.0$ & $50.00$ & $50.00$ & $2050.0$\\
3 & 2020-12-01T12:00 & $18.0$ & $-55.0$ & $0.0$ & $28.00$ & $28.00$ & $1350.0$\\
4 & 2020-12-03T23:05 & $-6.0$ & $-135.0$ & $0.0$ & $18.00$ & $18.00$ & $350.0$\\
5 & 2020-12-07T19:38 & $-22.0$ & $5.0$ & $0.0$ & $42.00$ & $42.00$ & $1380.0$\\
\bottomrule
\end{tabular}
\end{subtable}

\vspace*{0.08\linewidth}
    
\begin{tablenotes}
\textit{Note.} Each block shows, from left to right: CME number, time ($t_{0}$) of insertion of the CME at the Enlil inner boundary of $21\,R_{\odot}$ or 0.1~AU, latitude ($\theta$) and longitude ($\phi$) of the CME apex in Stonyhurst coordinates, tilt ($\gamma$) of the CME axis with respect to the solar equator (defined positive for counterclockwise rotations), semi-major ($\psi_{1}$) and semi-minor ($\psi_{2}$) axes of the CME cross-section (note that $\psi_{1} \neq \psi_{2}$ only in Run3), and CME radial speed at $21\,R_{\odot}$ ($V$).
\end{tablenotes}    

\end{table}

\begin{itemize}
\item \textbf{Run1:} For this run, we use CME input parameters derived in real time and stored at the Space Weather Database Of Notifications, Knowledge, Information (DONKI; \url{https://kauai.ccmc.gsfc.nasa.gov/DONKI/}), maintained by the Community Coordinated Modeling Center (CCMC) at the NASA Goddard Space Flight Center. The CMEs are analysed using the Space Weather Prediction Center CME Analysis Tool \citep[SWPC\_CAT;][]{millward2013}, in which a teardrop-shaped 3D structure is visually fitted to coronagraph data so that it best matches observations. Since DONKI runs are performed in real time, SOHO and STEREO-A data are used simultaneously only if available at the time of the analysis, otherwise single-spacecraft imagery is employed. We remark that STEREO-A data, if available, are in the so-called ``beacon'' \citep{biesecker2008} format, which is characterised by a lower processing level than science-level data. Links to the DONKI CME analysis results for each of the eruptions under study are provided in the Data Availability Statement section.
\item \textbf{Run2:} For this run, we derive CME input parameters by employing the cone model \citep{fisher1984,zhao2002}, consisting of a 3D conical base culminating in a spherical front, applied to simultaneous science-level data from SOHO and STEREO-A. In order to do so, we use the Graduated Cylindrical Shell \citep[GCS;][]{thernisien2006,thernisien2009} reconstruction technique, which consists of a parameterised shell intended to reproduce the 3D morphology of flux ropes (whose envelopes are closed coronal loops) that can be visually fitted to coronagraph images. Its free parameters are latitude ($\theta$), longitude ($\phi$), and height ($H$) of the CME apex, tilt ($\gamma$) of the CME axis, angular distance ($\alpha$) between the CME legs, and aspect ratio ($\kappa$). The cone model geometry is achieved by setting $\alpha$ to zero, and the value of $\gamma$ becomes irrelevant given the circular cross-section of the structure.  
\item \textbf{Run3:} For this run, we use the same methodology as in Run2, but allow the full range of GCS parameters to be freely adjusted until they best match simultaneous SOHO and STEREO-A coronagraph observations. The resulting CME morphology is reminiscent of a croissant, with both legs connected to the Sun and a toroidal axis. Furthermore, this is the only run in the mini-ensemble that allows for an elliptical cross-section of the resulting CME to be inserted in Enlil, often a more realistic representation \citep[e.g.,][]{cremades2004,krall2006}.
\item \textbf{Run4:}  The CME parameters for this run are obtained by fine-tuning the inputs of Run2 with the aim to match, to the best of our abilities, the observed CME-driven shock arrivals and arrival times at the six available in-situ locations described in Section~\ref{sec:data}. In other words, the CME input parameters are adjusted in order to minimise the rate of incorrect hit/miss predictions and the $\Delta{t}$ between observed and simulated arrival times derived from the other runs---in particular, Run2. This ``a posteriori'' run sets the benchmark in terms of what the present version of the coupled WSA--Enlil--SEPMOD models can achieve for the period under study. For example, shock--observer connectivities below $21\,R_{\odot}$ will be naturally missed because CMEs are not described beneath the Enlil inner boundary. Hence, this run will aid in discriminating between features that depend exclusively on the input parameters in the other three runs and those that cannot be reproduced with the current model setup even with optimal tuning.
\end{itemize}
For Run2 and Run3, the CME dimensions (with circular cross-section for Run2 and elliptical cross-section for Run3) are derived by ``slicing'' the image fitting-derived 3D structures at their widest point \citep[see][for additional notes and formulas on the model geometry]{thernisien2011}, and the time of injection at the Enlil inner boundary is derived by propagating the CMEs to $21.5\,R_{\odot}$ from their final observation in the combined C3+COR2 field of view under the assumption of constant speed (calculated between reconstructions performed at the times shown in Figure~\ref{fig:coronagraph} minus 60~minutes for CME1, CME3, and CME4, minus 30~minutes for the faster CME2 and CME5). In this respect, Run1 can be considered an actual forecast, Run2 and Run3 are hindcasts, and Run4 corresponds to post-event analysis. Links to all WSA--Enlil--SEPMOD runs can be found in the Data Availability Statement section.


\section{Simulation Results} \label{sec:results}

Simulation results for each of the six in-situ locations, including model--data comparisons, are shown in Appendix~\ref{app:runs} (Figures~\ref{fig:psp} to \ref{fig:solo}). Additionally, successive snapshots of the ecliptic plane for each of the four mini-ensemble runs are reported in Appendix~\ref{app:snaps} (Figures~\ref{fig:run1} to \ref{fig:run4}). According to in-situ measurements, CME1 impacted PSP, CME2 impacted PSP and STEREO-A, CME3 impacted STEREO-A, CME4 did not impact any observer, and CME5 impacted Earth and possibly Mars (more discussion of this aspect can be found in Section~\ref{subsec:arrtime}). Note that by ``impact'', we refer exclusively to the arrival of CME-driven shocks. None of the CMEs considered in this work impacted Bepi or SolO. SEPs from CME2/Flare1 were observed at all in-situ locations except for Bepi, whilst SEPs from CME5/Flare2 were detected at PSP, STEREO-A, Mars, and Earth, with possible effects at Bepi and SolO (more discussion of this aspect can be found in Section~\ref{subsec:seps}). Thus, particles from the first (second) SEP event spread over at least $240^{\circ}$ ($100^{\circ}$) in longitude. In the remainder of this section, we review and analyse the WSA--Enlil--SEPMOD simulation results in terms of three main aspects: CME-driven shock arrivals (Section~\ref{subsec:arrtime}), shock--observer connectivities (Section~\ref{subsec:connectivity}), and duration and peak fluxes of SEP events (Section~\ref{subsec:seps}).

\subsection{CME-driven Shock Arrival Time and Speed} \label{subsec:arrtime}

As described in Section~\ref{sec:setup}, the five CMEs modelled in this study are inserted in the Enlil heliospheric domain as hydrodynamic structures that are able to drive shocks in the heliosphere. Hence, the first step of our analysis is centred on evaluating whether a CME-driven shock is correctly predicted to impact each observing location by comparing the results of the different WSA-Enlil runs with the corresponding in-situ observations (shown in Appendix~\ref{app:runs}). As briefly mentioned above, we identify six total shock arrivals, which could be attributed to CME1 (at PSP), CME2 (at PSP and STEREO-A), CME3 (at STEREO-A), and CME5 (at Earth and Mars). Whilst the first five impacts present clear signatures of CME-driven shocks in in-situ measurements, the impact of a shock driven by CME5 at Mars is more ambiguous (see Figure~\ref{fig:mars}), in particular because of the lack of upstream  magnetic field data and the low cadence of plasma data.

Table~\ref{tab:arrival} shows contingency tables for each of the four mini-ensemble WSA--Enlil runs, where `hit' (H) means that a CME is correctly predicted to impact a target, `correct rejection' (R) indicates that a CME missed an observer in the model as well as in actual observations, and `false alarm' (F) denotes that a CME forecast to impact a certain location was not detected in situ. A fourth `miss' category is usually included in contingency tables, in this case signifying that a CME is incorrectly predicted to avoid a target. However, we have no instances of `miss' in our mini-ensemble. Overall, the contingency tables for the four runs are similar, as might be expected, but there are a few differences. In particular, the only instances of inaccurate forecasts in Table~\ref{tab:arrival} correspond to `false alarms' associated with CME1, CME2, and CME3. Note that `false alarms' at Mars for CME2 and CME3 are reported for all runs, including Run4 that was tuned to reflect in-situ measurements in terms of impacts and arrival times (see Section~\ref{sec:setup}). However, we remark that any clear deviation from the associated `ambient' run (without CMEs, also shown in Figures~\ref{fig:psp} to \ref{fig:solo}) is considered an impact here. Several are rather weak and could be difficult to recognise in the corresponding in-situ measurements. As for the other runs, Run1 was characterised by five `false alarms', whilst Run2 and Run3 each reported four `false alarms', indicating that these mini-ensemble members performed largely comparably in this regard.

\begin{table}[!t]
\centering
\caption{Contingency Tables for Each WSA--Enlil Simulation Run.}
\label{tab:arrival}

\begin{subtable}[b]{.4\linewidth}
\centering
\caption{Run1}
\begin{tabular}{c c c c c c}
\toprule
\#CME & 1 & 2 & 3 & 4 & 5 \\
\midrule
PSP & H & H & R & R & R \\
ST-A & \textcolor{purple}{F} & H & H & R & R \\
Mars & R & \textcolor{purple}{F} & \textcolor{purple}{F} & R & H \\
Earth & R & \textcolor{purple}{F} & \textcolor{purple}{F} & R & H \\
Bepi & R & R & R & R & R \\
SolO & R & R & R & R & R \\
\bottomrule
\end{tabular}
\end{subtable}
\begin{subtable}[b]{.4\linewidth}
\centering
\caption{Run2}
\begin{tabular}{c c c c c c}
\toprule
\#CME & 1 & 2 & 3 & 4 & 5 \\
\midrule
PSP & H  & H & R & R & R \\
ST-A & R & H & H & R & R \\
Mars & R & \textcolor{purple}{F} & \textcolor{purple}{F} & R & H \\
Earth & R & \textcolor{purple}{F} & \textcolor{purple}{F} & R & H \\
Bepi & R & R & R & R & R \\
SolO & R & R & R & R & R \\
\bottomrule
\end{tabular}
\end{subtable}

\begin{subtable}[b]{.4\linewidth}
\centering
\caption{Run3}
\begin{tabular}{c c c c c c}
\toprule
\#CME & 1 & 2 & 3 & 4 & 5 \\
\midrule
PSP & H  & H & R & R & R \\
ST-A & \textcolor{purple}{F} & H & H & R & R \\
Mars & R & \textcolor{purple}{F} & \textcolor{purple}{F} & R & H \\
Earth & R & R & \textcolor{purple}{F} & R & H \\
Bepi & R & R & R & R & R \\
SolO & R & R & R & R & R \\
\bottomrule
\end{tabular}
\end{subtable}
\begin{subtable}[b]{.4\linewidth}
\centering
\caption{Run4}
\begin{tabular}{c c c c c c}
\toprule
\#CME & 1 & 2 & 3 & 4 & 5 \\
\midrule
PSP & H  & H & R & R & R \\
ST-A & R & H & H & R & R \\
Mars & R & \textcolor{purple}{F} & \textcolor{purple}{F} & R & H \\
Earth & R & R & R & R & H \\
Bepi & R & R & R & R & R \\
SolO & R & R & R & R & R \\
\bottomrule
\end{tabular}
\end{subtable}

\vspace*{0.02\linewidth}
\begin{tablenotes}
\textit{Note.} H = hit; R = correct rejection; F = false alarm. An additional category usually employed in contingency tables is M = miss, with no occurrences in this particular case.
\end{tablenotes}
\end{table}

\begin{figure}[!t]
\centering
\includegraphics[width=.97\linewidth]{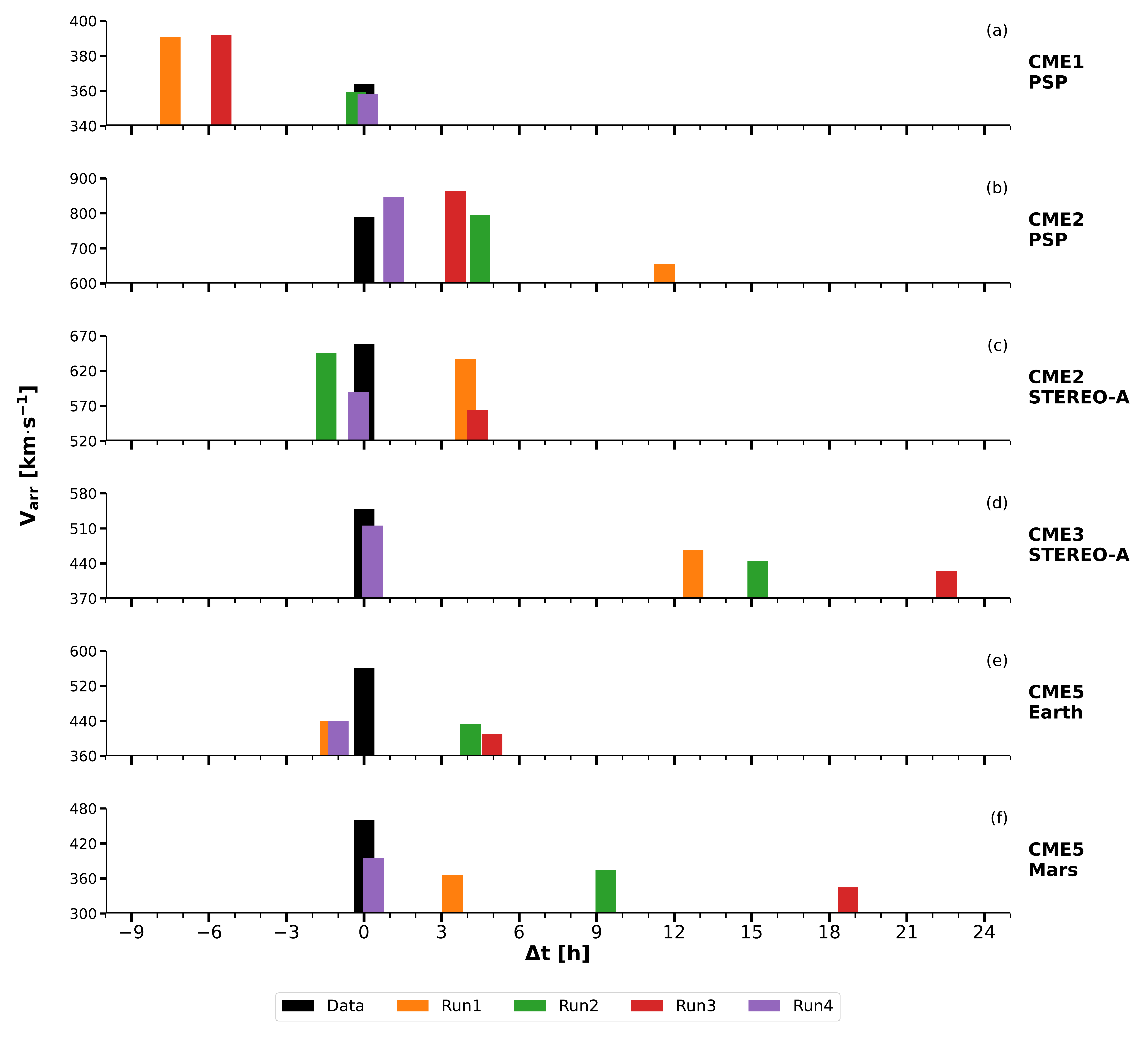}
\caption{Difference between simulated and in-situ arrival times (${\Delta}t$) versus arrival speeds ($V_\mathrm{arr}$) shown for each location where a CME-driven shock was detected and for each of the WSA--Enlil model runs.}
\label{fig:arrtimes}
\end{figure}

An overview of the differences between the observed and predicted CME-driven shock arrival times and speeds for the six `hit' instances for the four runs in the mini-ensemble is shown in Figure~\ref{fig:arrtimes}. Regarding the arrival times, apart from Run4, in which errors were minimised by design, there is no run that consistently performed best at all locations. Nevertheless, there are differences in the cumulative error over the six impacts: 40.5~hours for Run1, 31.4~hours for Run2, 59.6~hours for Run3, and 3.2~hours for Run4. Thus, Run3 performed worst overall, which may seem surprising since the ellipsoidal morphology included in this run was intended to correspond to a more realistic CME representation (see Section~\ref{sec:setup}). The poorer performance of Run1 compared with Run2 may arise because two remote-sensing viewpoints were not always available when the CME input parameters for Run1 were issued in real time (see Section~\ref{sec:setup}) and lower-quality STEREO-A beacon images were used. Nevertheless, the difference between Run1 and Run2 in $\Delta t$ averages to only ${\sim}1.5$~hours per event (${\sim}9$~hours over six CME impacts), suggesting that remote-sensing data available in real time may be sufficient for the purpose of CME arrival time forecasting with Enlil---at least with the observing spacecraft configuration explored in this work, i.e., L1+L5.

It is also important to critically consider these results in relation to the background solar wind modelled by WSA--Enlil through which the CMEs are propagating, which is reflected in both the modelled arrival times and speeds. One clear example of this is CME5 at Earth (Figure~\ref{fig:arrtimes}(e)), where Run1 achieved a more accurate arrival time than Run2 and Run3; nevertheless, the Enlil ambient solar wind preceding the event was ${>}100$~km$\cdot$s$^{-1}$ slower than observed, indicating that the ``true'' CME speed was likely closer to that set for the worse performing runs. A slower background wind also affects CME drag and deceleration, resulting in this case in all the simulated arrival speeds being significantly ($\Delta V > 100$~km$\cdot$s$^{-1}$) lower than observed. This aspect may also explain why the general trend shown in Figure~\ref{fig:arrtimes} is that all runs tend to estimate a later CME arrival time: Most impacts are preceded by slower ambient speeds modelled by Enlil (see Appendix~\ref{app:runs}), suggesting that this is due to an intrinsic underestimation of the background solar wind speed for this particular time period rather than, e.g., systematic human error when evaluating CME input parameters. In fact, 13 (72\%) of the 18 arrival times for the first three runs (i.e., excluding Run4) are characterised by $\Delta t \geq 3$~hours, only 2 (11\%) feature $\Delta t \leq {-}3$~hours, and 3 (17\%) result in arrival times within 3~hours of the observed ones. At the same time, 14 (58\%) of the 24 total shock arrival speeds are lower than observed by at least 50~km$\cdot$s$^{-1}$, and this is true even for Run4 in half of the cases (CME2/STEREO-A, CME5/Earth, and CME5/Mars). In contrast, only 2 (8\%) modelled speeds are faster than observed by at least 50~km$\cdot$s$^{-1}$, whilst the remaining 8 (34\%) impacts are within 50~km$\cdot$s$^{-1}$ of the observed arrival speeds.

\subsection{Shock--Observer Connectivity} \label{subsec:connectivity}

\begin{figure}[p]
\centering
\hspace*{-0.06\linewidth}
\includegraphics[width=1.12\linewidth]{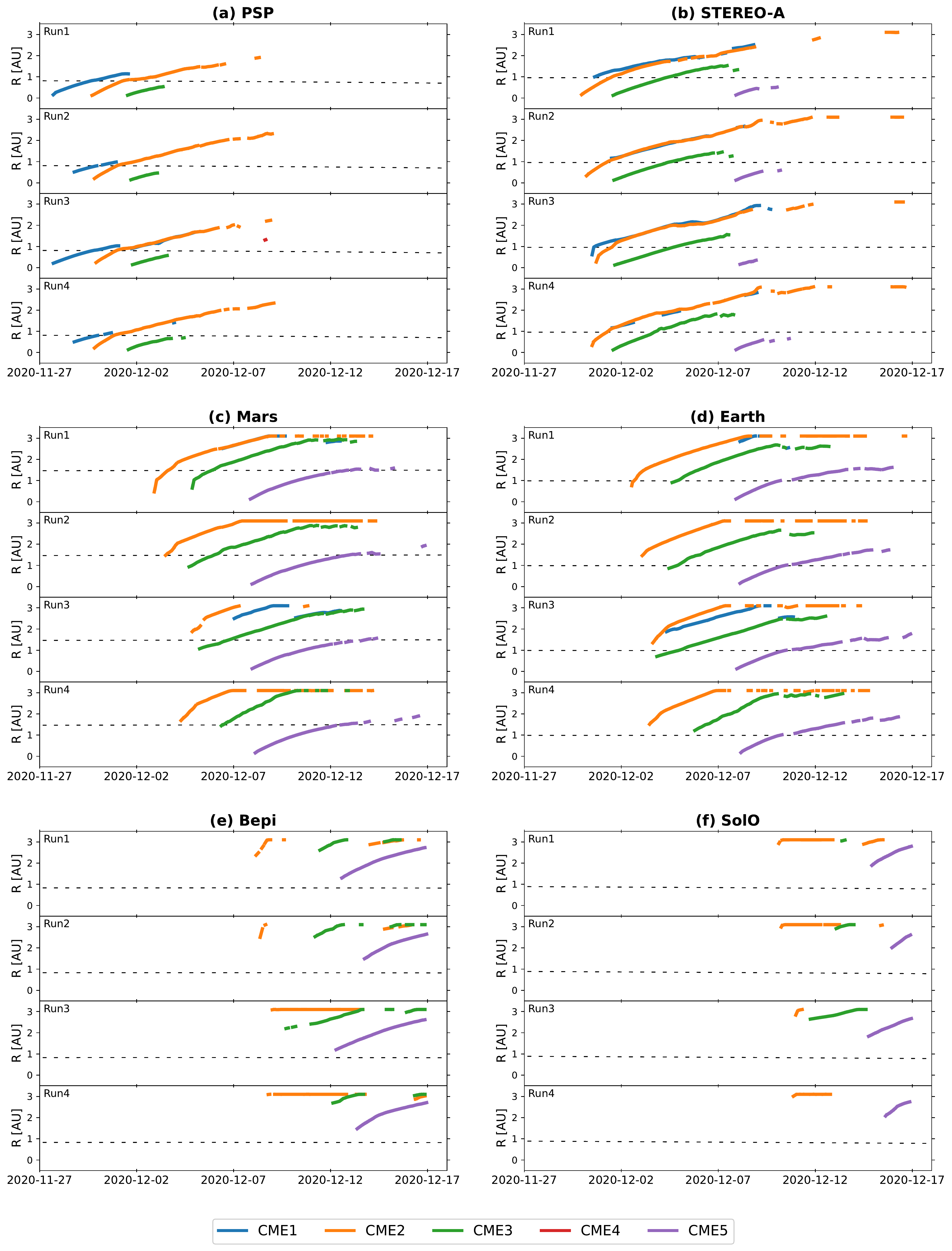}
\caption{Shock--observer connectivity shown in terms of the shock radial distance along the observer-connected field line. Connectivity is shown for the five CMEs in this study with respect to each of the four mini-ensemble simulation runs, at (a) PSP, (b) STEREO-A, (c) Mars, (d) Earth, (e) Bepi, and (f) SolO. The dashed line in each panel indicates the time-dependent heliocentric distance of the corresponding observer.}
\label{fig:connectivity}
\end{figure}

Another interesting aspect to analyse is the connectivity of the CME shocks to the six in-situ locations with respect to each WSA--Enlil simulation run. This is particularly important for SEPMOD, since this model requires connection to a shock for accelerated particles to be detected at a particular location. In Enlil, shocks are defined as ${\geq}20$~km$\cdot$s$^{-1}$ increases in solar wind speed with respect to the corresponding ambient simulation (without CMEs), and the shock speed is derived separately as a time-derivative of the shock position. Connectivity also directly affects the SEP intensity--time profiles modelled by SEPMOD. Figure~\ref{fig:connectivity} shows the intervals of shock--observer connectivity for each CME in terms of the shock radial distance along the field line connected to the observer. In Enlil simulations, a shock may become magnetically connected to an observer at any point within the simulation domain, which in our runs corresponds to heliocentric distances between 0.1 and 3~AU. This means that a shock may first connect to an observer ``from behind'', when the shock is already beyond the radial distance of the observer (indicated by the dashed lines in Figure~\ref{fig:connectivity}). Several examples of connection from behind are evident in the figure.

According to the modelling results shown in Figure~\ref{fig:connectivity}, the shock driven by CME2 (corresponding to Flare1) connects with every observer at some point throughout the temporal domain for each simulation run. However, connectivity is established when the shock is still sunward of the spacecraft only at PSP and STEREO-A, while Mars and Earth connect with the shock when it is at or beyond their respective heliocentric distances in all cases except for Run1. Bepi and SolO become magnetically linked to the CME2 shock only when it is well beyond 2~AU, approximately 10~days after the eruption of Flare1. These results are consistent with the locations of the various observers with respect to Flare1 shown in Figure~\ref{fig:orbits}a. The second SEP-rich event, i.e.\ CME5 (associated with Flare2), becomes magnetically connected to all observers except PSP. Connectivity is achieved ``from the front'' at STEREO-A, Mars, and Earth, and ``from behind'' at Bepi and SolO. Again, these results are generally as expected considering the spacecraft configuration relative to the location of Flare2 shown in Figure~\ref{fig:orbits}b, with STEREO-A, Earth, and Mars favourably positioned for early connection to the shock, whilst Bepi and SolO behind the west limb connect to the shock once it has propagated well beyond 1~AU. The absence of connectivity at PSP can be accounted for because the modelled CME when sunward of the spacecraft is not sufficiently extended in longitude to reach the field line connecting to PSP (see the figures in Appendix~\ref{app:snaps}), and PSP is not suitably located to be connected to the shock from behind. The connectivities for CME1, CME3, and CME4 are more dependent on the specific simulation run. CME1 becomes connected to PSP and STEREO-A in all runs, but connectivity with Earth and Mars is achieved only in Run1 and Run3. CME3 connects to all observers in Run1, Run2, and Run3, but misses SolO in Run4. Finally, CME4 becomes briefly (for ${\sim}1$~hour) connected to PSP only in Run3.

Overall, the connectivity curves shown in Figure~\ref{fig:connectivity} are largely similar for each observer across the four simulation runs, apart from the few differences discussed above. The major differences between the runs at any given observer are in the times when connectivity to a shock is first achieved, with spreads ranging from a few hours to ${\sim}2.5$~days. This indicates that the connectivity to a shock modelled by Enlil is more sensitive to the CME input parameters than the shock arrival time, which only shows a range of less than 24 hours across the four runs (see Section~\ref{subsec:arrtime}). As an example, for CME1 at PSP, the spread in arrival times for the CME-driven shock is ${\sim}8$~hours, whereas the spread in magnetic connection onset times is about ${\sim}27$~hours. This spread in the first connection times significantly affects the SEP intensity--time profiles modelled by SEPMOD for each run, as explored in greater detail in the next section. Moreover, we highlight the expected sources of error related to the employed modelling methods. First of all, connectivity cannot be modelled below 0.1~AU---the height of Enlil's inner boundary at which the modelled CMEs are inserted. Hence, early CME acceleration and shock formation cannot be realistically described. In addition, the lack of CME internal magnetic fields is expected to yield intrinsic uncertainties in the resulting interplanetary magnetic field preconditioning and, thus, in the modelled connectivities. This may hold true especially in the case of CME2, which was launched shortly after and on a similar trajectory to CME1.

\subsection{SEP Event Duration and Flux} \label{subsec:seps}

The figures in Appendix~\ref{app:runs} show the SEP proton intensity--time profiles generated by SEPMOD at three sample energies together with observations from each spacecraft again in three proton energy ranges, which vary between instruments. In order for particles to be detected by a spacecraft, SEPMOD requires that the observer encounters a magnetic field line that at some time was connected with the shock and therefore populated with energetic particles injected by the shock. The intensity and energy spectrum of these particles depend on the shock compression ratio and velocity jump, respectively, at the point where the field line intersects the shock. The figures in Appendix~\ref{app:runs} show that the major SEP enhancements observed at the various locations were associated with CME2/Flare1 and CME5/Flare2. All observers except Bepi detected the first SEP event, whilst the second SEP event was detected at least at PSP, STEREO-A, Mars, and Earth. Bepi and SolO recorded low-energy SEP enhancements following the eruption of Flare2 (see panel (f) of Figures~\ref{fig:bepi} and \ref{fig:solo}), but it is unclear whether these were associated with this event or with the passage of a stream interaction region \citep[SIR; e.g.,][]{richardson2018a} that appears to corotate from Bepi to SolO. Although plasma observations are incomplete, the Enlil results at Bepi show an SIR on December~8, associated with an observed magnetic field enhancement and a sector crossing, that is also evident at SolO on December~10.  An additional complication is that Bepi and SolO detected a minor SEP event with onset around 00:00~UT on 2020 December~11. This was likely related with a CME eruption late on December~10 from behind the west (east) limb as seen from Earth (STEREO-A) that was well positioned for Bepi and SolO but was outside our CME analysis interval. There is also a weak SEP event associated with CME1 observed at PSP (see Figure~\ref{fig:psp}). 

\begin{table}[!t]
\centering
\caption{Contingency Tables for Each WSA--Enlil--SEPMOD Simulation Run.}
\label{tab:sep}

\begin{subtable}[b]{.4\linewidth}
\centering
\caption{Run1}
\begin{tabular}{c c c}
\toprule
\#CME &  2 & 5 \\ 
& lo-mi-hi & lo-mi-hi \\
\midrule
PSP & H H H & \textcolor{purple}{M} \textcolor{purple}{M} \textcolor{purple}{M} \\
ST-A & H H H & H H h \\
Mars & H H H & H H \textcolor{purple}{F} \\
Earth & H H h & H H H \\
Bepi & R R R & ? ? ? \\
SolO & \textcolor{purple}{M} \textcolor{purple}{M} \textcolor{purple}{M} & ? ? ? \\
\bottomrule
\end{tabular}
\end{subtable}
\begin{subtable}[b]{.4\linewidth}
\centering
\caption{Run2}
\begin{tabular}{c c c}
\toprule
\#CME &  2 & 5 \\ 
& lo-mi-hi & lo-mi-hi \\
\midrule
PSP & H H H & \textcolor{purple}{M} \textcolor{purple}{M} \textcolor{purple}{M} \\
ST-A & H H H & H H H \\
Mars & H H H & H H \textcolor{purple}{F} \\
Earth & H H h & H H H \\
Bepi & R R R & ? ? ? \\
SolO & \textcolor{purple}{M} \textcolor{purple}{M} \textcolor{purple}{M} & ? ? ? \\
\bottomrule
\end{tabular}
\end{subtable}

\begin{subtable}[b]{.4\linewidth}
\centering
\caption{Run3}
\begin{tabular}{c c c}
\toprule
\#CME &  2 & 5 \\ 
& lo-mi-hi & lo-mi-hi \\
\midrule
PSP & H H H & \textcolor{purple}{M} \textcolor{purple}{M} \textcolor{purple}{M} \\
ST-A & H H H & \textcolor{purple}{M} \textcolor{purple}{M} \textcolor{purple}{M} \\
Mars & H H H & H H \textcolor{purple}{F} \\
Earth & H H h & H H H \\
Bepi & R R R & ? ? ? \\
SolO & \textcolor{purple}{M} \textcolor{purple}{M} \textcolor{purple}{M} & ? ? ? \\
\bottomrule
\end{tabular}
\end{subtable}
\begin{subtable}[b]{.4\linewidth}
\centering
\caption{Run4}
\begin{tabular}{c c c}
\toprule
\#CME &  2 & 5 \\ 
& lo-mi-hi & lo-mi-hi \\
\midrule
PSP & H H H & \textcolor{purple}{M} \textcolor{purple}{M} \textcolor{purple}{M} \\
ST-A & H H H & H H H \\
Mars & h H H & H H \textcolor{purple}{F} \\
Earth & H H \textcolor{purple}{M} & H H H \\
Bepi & R R R & ? ? ? \\
SolO & \textcolor{purple}{M} \textcolor{purple}{M} \textcolor{purple}{M} & ? ? ? \\
\bottomrule
\end{tabular}
\end{subtable}

\vspace*{0.02\linewidth}
\begin{tablenotes}
\textit{Note.} H = hit; R = correct rejection; M = miss, F = false alarm. A `h' is used for H instances for which the modelled SEP fluxes peaked barely above the corresponding instrument's background. The three values for each CME (`lo', `mi', and `hi') refer to the SEP low, mid, and high energies shown respectively in panels  (f), (g), and (h) of Figures~\ref{fig:psp} to \ref{fig:solo}.
\end{tablenotes}
\end{table}

Table~\ref{tab:sep} shows contingency tables for each of the SEPMOD runs in relation to the SEP events driven by CME2 and CME5, in a similar fashion to Table~\ref{tab:arrival}. For CME1 (not shown in Table~\ref{tab:sep}), all runs predict a lower-energy particle (${\sim}5$~MeV) enhancement that far exceeds the observed intensities that rise just above the instrument noise level, whilst protons with energies ${\sim}25$~MeV are predicted only in Run1 and Run3. SEPs from CME2 are predicted to reach all observers, including Bepi, which in reality missed the event. However, the predicted arrival at Bepi and SolO is approximately 10~days after the eruption and occurs because of connection to the shock well beyond the spacecraft (Figure~\ref{fig:connectivity}(e--f)). In practical terms, these predictions can be considered as `misses' (or `correct rejections'). Due to the temporal proximity of CME1, CME2, and CME3, potentially the particle fluxes at some locations might have contributions from all three CMEs. Inspection of the SEPMOD predicted fluxes for each CME separately suggests that these contributions were likely to be negligible compared to that from the larger and more energetic CME2. An exception is Run3 at Earth and Mars, where CME3 makes a significant contribution to the resulting predicted SEP profile, though it is not clear from the observed fluxes that there was an actual contribution from CME3. The observed SEPs from CME5 were not predicted at PSP, or at STEREO-A for Run3 only, and were predicted to reach SolO and Bepi ${\sim}$5--8~days later, after the small particle enhancement (possibly associated with an SIR) discussed above (note that, because of the dubious association of the weak SEP event at Bepi and SolO, the corresponding values in Table~\ref{tab:sep} are indicated with a question mark). All remaining SEPMOD SEP predictions for both CME2 and CME5 can be classified as `hits', whilst predictions for CME4 are `correct rejections' at all locations. Overall, all runs performed largely comparably, with a slightly worse performance achieved by Run3. We emphasise that the contingency tables in Table~\ref{tab:sep} are based entirely on the model--data comparisons shown in Figures~\ref{fig:psp} to \ref{fig:solo}, hence they are dependent on factors such as the instrumental background noise. For example, in the case of the SEP event driven by CME2, Run1, Run2, and Run3 predict enhancements that are barely above the noise level (`h'), whilst Run4 predicts an enhancement that is just below the background (`M'). In reality, the obseved SEP event itself was not significantly above background levels, and all runs predicted fluxes within a factor 5 of the measurements.

\begin{figure}[t!]
\centering
\hspace*{-0.06\linewidth}
\includegraphics[width=1.12\linewidth]{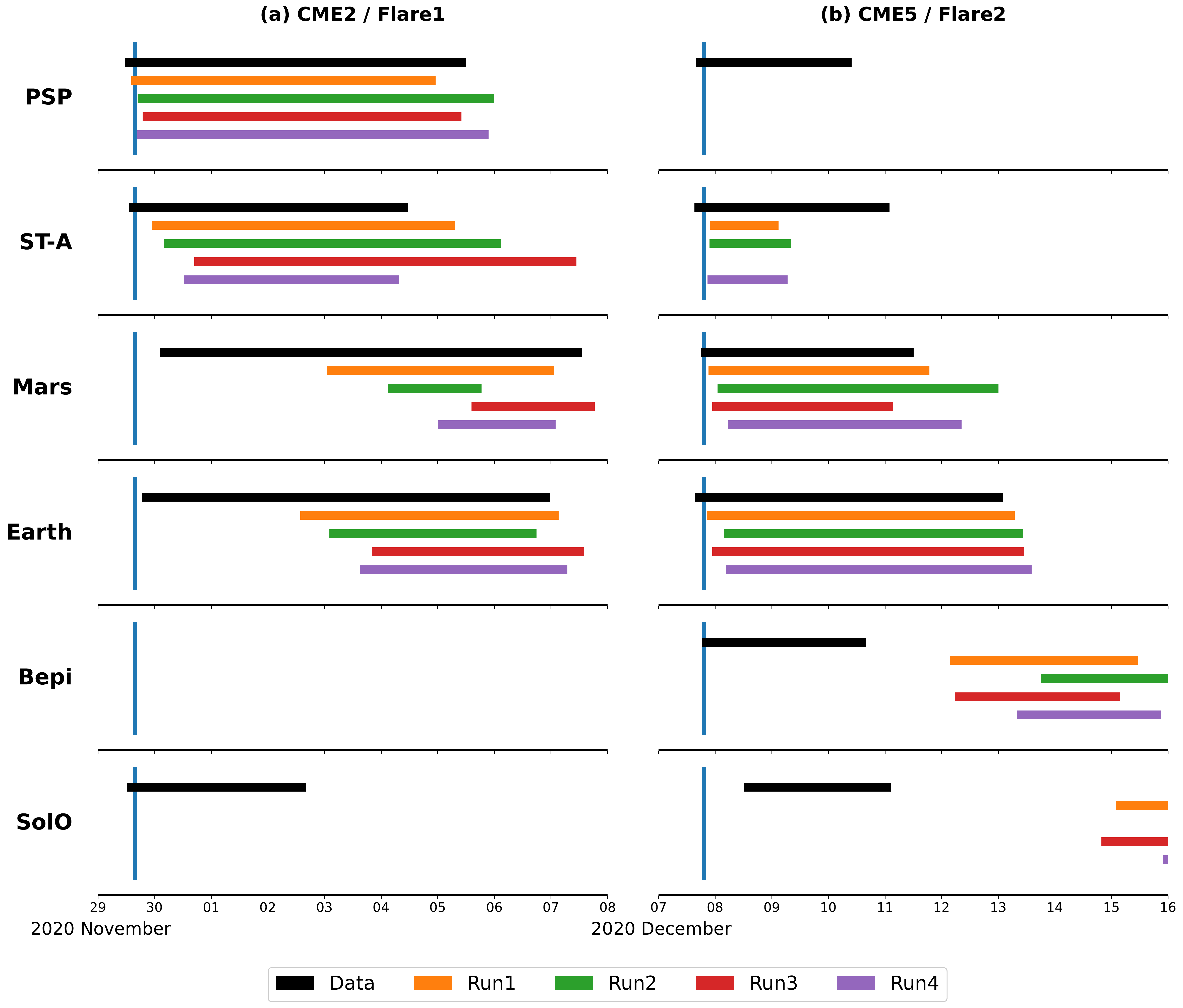}
\caption{Temporal extent of the SEP enhancements associated with (a) CME2/Flare1 and (b) CME5/Flare2, at the six in-situ locations, showing data (black) versus the mini-ensemble runs (colours). These are based on ``mid-energy particles'' (panel (g) in Figures~\ref{fig:psp} to \ref{fig:solo}) measurements everywhere apart from Bepi and SolO for CME5/Flare2, where the event was possibly observed only at lower energies (panel (f) in Figures~\ref{fig:bepi} and \ref{fig:solo}). The blue vertical lines mark the injection times of the corresponding CMEs into the Enlil heliospheric domain, which can be considered uniform across the four mini-ensemble runs in the scale shown (see Table~\ref{tab:cme_list}).}
\label{fig:sepextent}
\end{figure}

The temporal extents of the SEP enhancements associated with CME2/Flare1 and CME5/Flare2 modelled by SEPMOD are compared with those observed in situ at ${\sim}25$~MeV (panel (g) in Figures~\ref{fig:psp} to \ref{fig:solo}) in Figure~\ref{fig:sepextent}. In line with the results shown in Table~\ref{tab:sep}, the occurrence of an SEP event is successfully predicted at most locations. However, the predicted onset times of these events are later than observed in all cases, with delays ranging from a few hours up to several days. The different event onset times for the four simulation runs reflect the differences in shock--observer connectivity as discussed in Section~\ref{subsec:connectivity}. In at least half of the cases, the observed particle onset times occur before the CME injection time into Enlil (represented by a vertical line in each panel), indicating that magnetic connectivity was likely established when the shock was still below 21.5\,$R_{\odot}$. In such cases, it is impossible to predict the correct SEP onset time with the heliospheric modelling setup used in this work. SEPMOD best predicts the temporal durations of the SEP enhancements at PSP and STEREO-A for CME2, and at Mars and Earth (and to some extent at STEREO-A) for CME5. These locations were closest in terms of longitude to the respective CME source regions. Note that here we consider the results for CME5/STEREO-A to be more successful than those for CME2/Earth and CME2/Mars. This is because, in the former case, approximately the first half of the event is captured, i.e.\ close to its peak, whilst capturing the second half of an event corresponds to its declining phase, when particle fluxes are generally lower. The least successful predictions are for Bepi and SolO, for which the sources of both CMEs were at or behind the solar limb from their viewpoints, and for PSP in the case of CME5/Flare2. Thus, these results indicate that the SEP prediction performance declines with longitudinal distance from the eruption site. In Figure~\ref{fig:sepextent}(b), the small particle enhancements observed at Bepi and SolO have been associated with CME5 even if their origin is not fully clear and they may be related to an SIR as discussed above. In any case, the SEPMOD results at these two locations suggest that the particle events would commence significantly later than observed (almost a `miss'), or that they were simply a `false alarm'. Finally, there are no significant differences across the four mini-ensemble members, even including Run4 that was designed to match the observed CME arrival times (see Section~\ref{sec:setup}). This suggests that an improved CME-shock arrival forecast will not automatically result in a better SEP forecast. The performance is only slightly worse for Run3, which missed the SEP event driven by CME5 at STEREO-A and predicted a larger contribution from CME3 in the SEP event driven by CME2.

\begin{figure}[ht]
\centering
\hspace*{-0.06\linewidth}
\includegraphics[width=1.12\linewidth]{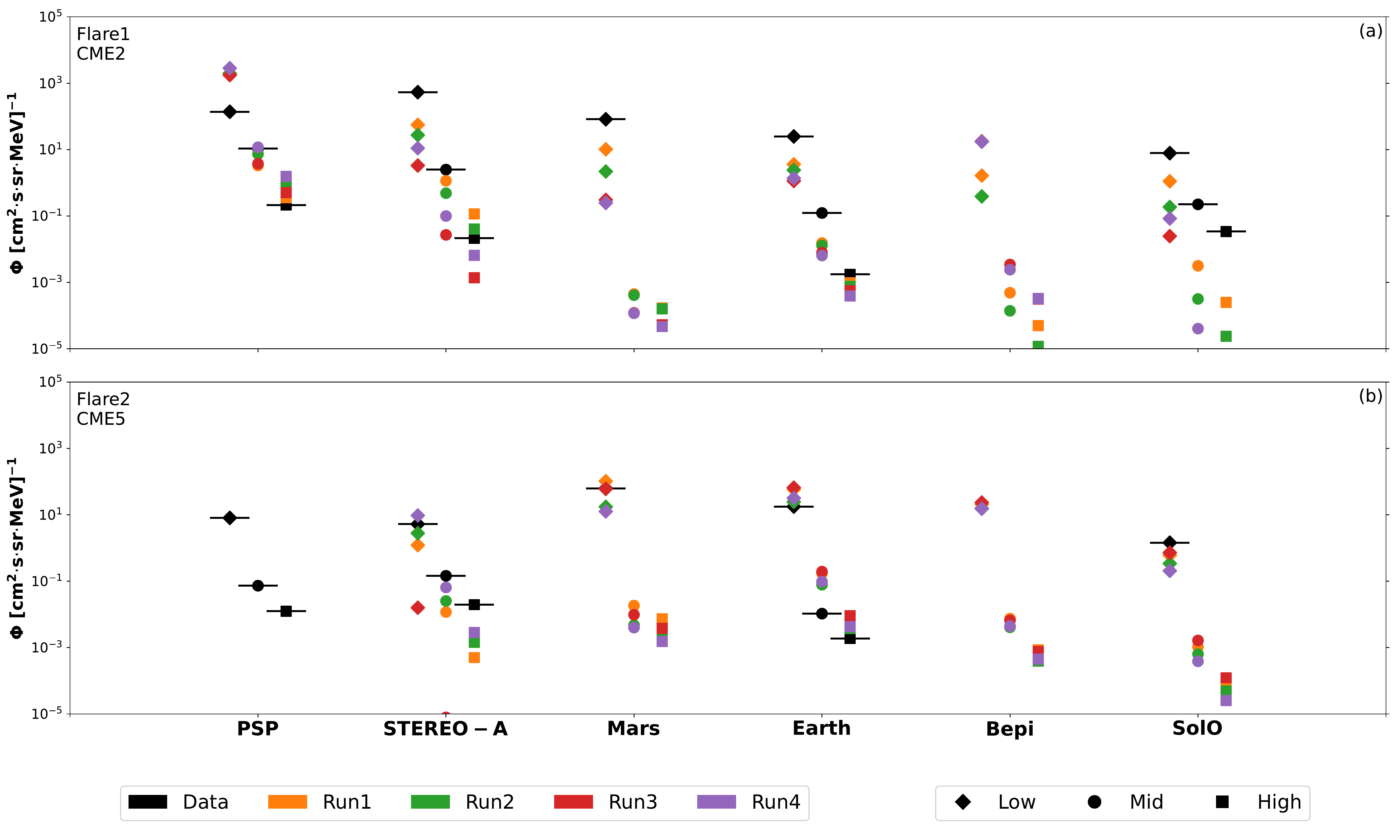}
\caption{Maximum SEP flux associated with (a) CME2 or Flare1 and (b) CME5 or Flare2, showing data (black) versus the mini-ensemble runs (colours). The three energy levels (`low', `mid', and `high') correspond to those reported in panels (f), (g), and (h) of Figures~\ref{fig:psp} to \ref{fig:solo}, respectively. The `data' markers are accompanied by a horizontal bar for clarity.}
\label{fig:fluxes}
\end{figure}

The maximum SEP fluxes predicted by SEPMOD in each run at every observing location are compared with the maximum fluxes observed in situ in Figure~\ref{fig:fluxes}. The energy ranges, from left to right, correspond to those shown in panels (f), (g), and (h) of Figures~\ref{fig:psp} to \ref{fig:solo}. Since the energy ranges differ for each instrument, the fluxes  should not be compared directly across the six observers. It is clear that, apart from a few cases, SEPMOD generally tends to underestimate the maximum fluxes. It is possible that the lack of a realistic shock description below 21.5\,$R_{\odot}$, where CMEs tend to travel faster and accelerate higher-energy particles (before they are slowed down via solar wind drag), together with delayed shock--observer connectivities, result in an underestimation of particle acceleration and, thus, of peak fluxes. Of the 84 (21$\times$4) instances of available data--model comparisons shown in Figure~\ref{fig:fluxes}, 41 (49\%) cases underestimate and 10 (12\%) overestimate the peak fluxes by a factor ${\geq}$5, whilst 33 (39\%) lie within a factor 5 of the measurements (in either direction). If we consider a factor 3, these values become, respectively, 49 (58\%), 15 (18\%), and 20 (24\%). We note, however, that in the case of CME2 at PSP, fluxes at low energies seem to be greatly overestimated, but this is not likely to be the case since PSP data are accompanied by a large data gap centred on the CME2 shock arrival, where the ${\sim}$4-MeV particle intensity tends to peak. If we consider only the best-connected observers---i.e., PSP (excluding the low energies) and STEREO-A for CME2, as well as STEREO-A, Earth, and Mars for CME5---the percentages of underestimated--overestimated--acceptable fluxes become 33\%--13\%--54\% for a factor 5 and 44\%--23\%--33\% for a factor 3. This suggests that SEP peak fluxes tend to be better modelled for well-connected observers. Again, there is no mini-ensemble member that performed consistently better than the others, the only significant outlier being Run3 that predicted exceptionally low SEP fluxes at STEREO-A for CME5 compared to the other runs (which translated into a `miss' in Table~\ref{tab:sep}). In particular, we note that Run4 often does not provide the best results, again showing that improving CME arrival predictions may not lead to improved SEP predictions.


\section{Discussion} \label{sec:discussion}

The results presented in Section~\ref{sec:results} suggest a number of discussion points relevant to both CME and SEP predictions. The four mini-ensemble runs employed in this study provide an opportunity to analyse and review simulation outcomes based on using real-time forecasts, science-level hindcasts (assuming circular or elliptical cross-section CMEs), and post-event analysis. The input parameters used for each run were discussed in Section~\ref{sec:setup}.

First of all, all the mini-ensemble members correctly reported all the CME-driven shock impacts that were detected in situ (i.e., no CME arrival was missed). All the runs, however, were characterised by a few false alarms, albeit with different degrees of importance---most were only weak flank encounters. The worst-performing simulation run in terms of shock arrival time was Run3, using hindcasts performed with elliptical cross-section CMEs. This is rather unexpected, since a non-circular CME cross section is intended to provide a more realistic CME morphology. We note however that \citet{amerstorfer2021} reached similar conclusions when deriving input parameters for STEREO heliospheric imager-based CME propagation and forecasting models. They reported that simply assuming the CME half-width to lie within a certain range led to more accurate predictions than using GCS-derived values. The authors speculated that this may be due to CME deflections and/or rotations taking place beyond the coronagraphs' field of view. We also conjecture that the greater number of free parameters available when performing GCS reconstructions with a croissant morphology may lead to additional uncertainties and thus less precise estimates of the CME inputs for Run3. It is possible that more ``elongated'', non-circular, CME cross-sections provide more realistic, and better performing, morphologies for CMEs that include magnetic flux ropes, but if CME internal magnetic fields are not considered, as in this study, then simply assuming a circular cross-section seems to yield better results. We also note that Run2 (using hindcasts with a circular CME cross-sections) performed slightly better than Run1 (using parameters from real-time forecasts). The poorer performance of Run1 may be due to the lack of data from one coronagraph (either SOHO or STEREO-A) at the time when the prediction was issued (which is not always reported, hence we cannot draw solid conclusions) and/or to the fact that real-time STEREO-A imagery is only available as beacon data. \citet{bauer2021} showed that beacon data tend to perform worse (albeit not dramatically worse, as is the case in this study) than science data for space weather forecasting based on the heliospheric imaging cameras onboard STEREO. This work employs coronagraph rather than heliospheric imagery, but it is reasonable to assume that similar issues withstand. 

Though not considered in the mini-ensemble used in this work, where just the CME parameters are varied, CME arrival time predictions are also inevitably affected by the solar wind background that is modelled by Enlil using a model-generated coronal magnetic field input---in this case, from the WSA model. Since WSA uses maps of the photospheric magnetic field as input, differences in the input maps from specific ground- or space-based observatories and their calibration are expected to affect the modelled background solar wind. Therefore, an ensemble approach to modelling the background solar wind using different magnetic field maps combined with ensemble CME modelling may lead to more robust CME arrival forecasts. The influence of different input magnetic field maps in the context of global MHD simulations of the heliosphere has been explored by \citet{jin2022}, whilst more information on the impact of inner heliospheric boundary conditions generated with different models can be found in \citet{gonzi2021}. Additionally, it is important to consider the uncertainties regarding the predicted shock arrival time that are intrinsic to the choice of CME propagation model---in this case, Enlil. These have been explored in several studies, either within a single model \citep[e.g.,][]{calogovic2021,dumbovic2021,wold2018} or by comparing results from different models \citep[e.g.,][see also CCMC's CME Scoreboard: \url{https://ccmc.gsfc.nasa.gov/scoreboards/cme}]{paouris2021,riley2018}. At the time of the events studied, STEREO-A was approximately at the L5 point. The fact that all the errors in the CME arrival time results described here were similar to the ``typical errors'' in arrival time obtained in previous studies (see Introduction), and no `misses' occurred, suggests that remote-sensing monitoring from the Lagrange L1 and L5 points can provide adequate constraints on the propagation direction of CMEs.

Considering shock--observer connectivities, we found that the differences in CME parameters amongst the four runs had a greater effect on the time at which a location becomes first connected to the associated shock (differences across the mini-ensemble members exceeded 2~days) than on the shock arrival time (differences of less than 1~day). The connectivity of the various shocks to each observer was largely similar across the four simulation runs. The variation in the connectivity onset and duration suggests that an ensemble approach may be able to provide a more robust range of possible outcomes. However, two significant limitations of the connectivity modelling in this study are that CMEs are introduced into Enlil at its heliospheric inner boundary ($21.5\,R_{\odot}$ or 0.1~AU), resulting in delayed shock--observer connectivity onsets and the absence of a realistic CME acceleration profile, and that CMEs lack an internal magnetic field, resulting in a less realistic description of interplanetary magnetic field preconditioning. We also note that, in several cases, magnetic connection to the shock was established ``from behind'', when the shock front was beyond the observer's heliocentric distance. These instances take place in the model when an observer is well-separated in longitude from the shock nose, and tend to result in even longer delays before connection to the shock occurs. Nevertheless, as noted by \citet{luhmann2018}, in SEPMOD, shock connections far from the Sun provide considerably reduced SEP contributions compared to those near the Sun, since CME-driven shocks tend to weaken with heliocentric distance. \citet{bain2016} performed a multi-point study of the Enlil shock--observer connectivity for two periods characterised by several successive eruptions in a similar fashion to the ones analysed here, and compared results with SEP measurements at Earth, STEREO-A, and STEREO-B. They found that about two-thirds of the shock connections identified in the simulations occurred within a few hours of an observed SEP event, but also that a number of initial connections were missed due to the modelled shocks being initiated at $21.5\,R_{\odot}$. Our results support these conclusions.

Regarding SEP predictions, SEPMOD requires the existence of a shock--observer magnetic field connection and the predicted SEP properties depend on the shock parameters at the connection point. Here we consider the three main areas that were identified by \citet{bain2021} to be of high importance (see also the Introduction), i.e.\ SEP onset time, event duration, and peak fluxes. We found that the durations and peak fluxes of SEP events are generally well reproduced, but these predictions tend to become less accurate with increasing longitudinal distance from the CME eruption site. A likely reason in this respect is that SEPMOD assumes scatter-free propagation along well-defined field lines that are connected to a shock, whilst additional particle transport effects \citep[e.g.,][]{qin2004} are neglected in the model. These include, e.g., diffusive \citep[e.g.,][]{zank2015} and perpendicular \citep[e.g.,][]{zhang2003} transport, which are expected to play a more significant role in widespread SEP events \citep[][found that possible reasons for SEPs spreading broadly in longitude are cross-field transport in the interplanetary medium and/or particularly extended particle injection regions close to the Sun]{dresing2014}. In fact, these additional transport mechanisms may explain the two major `miss' instances in the modelled SEP events, i.e.\ those associated with CME2 at SolO and CME5 at PSP, which are characterised by observers that are so far from the eruption site that connectivity to the shock is either never achieved, or achieved only much later from behind. Hence, it is natural to assume that SEPMOD results and predictions are more accurate for well-connected observers, where scatter-free transport plays a major role. 

Interestingly, SEPMOD results from Run4 were not significantly better than those from the other runs, even though the CME shock arrival times were matched with observations by design. This may be due to the fact that, in order to reduce the amount of `false alarms' for shock arrivals from the previous runs (see Section~\ref{subsec:arrtime}), the sizes of some CMEs were somewhat diminished, thus delaying the time at which a well-separated observer becomes first connected to a shock. The least successful predictions of the model runs are for the SEP event onset times, which were delayed a few hours even in the best-connected cases. The SEP events at Bepi and SolO were predicted to commence several days later than the corresponding CME/flare eruption time, due to connectivities reached from behind, and particles at PSP from CME5 were entirely missed. However, as discussed in relation to shock connectivity, this is entirely to be expected because shocks are not modelled below $21.5\,R_{\odot}$, whereas it is well known that shocks associated with fast CMEs can often form in the low corona, and some can even be observed in coronagraph imagery \citep[e.g.,][]{vourlidas2013}. Furthermore, SEP events can clearly commence when shocks are much lower in the corona, as is evident from the results in this work and many other studies \citep[e.g.,][]{agueda2014,gopalswamy2013}. Additionally, it is possible that a SEP event has contributions from both a flare source that remains rooted in the corona and a CME-driven shock within the same eruptive event \citep[e.g.,][]{cane2010}. Thus, as noted earlier, the modelling setup used here may not be the best option to predict SEP onset times. In addition, obtaining the CME parameters usually requires the CME to be detected and tracked in coronagraph images, at which point a SEP event may have already commenced, even if the images were available in real time. It is also possible to initiate a WSA--Enlil simulation based on real-time EUV imagery only, but the CME input parameters will likely be significantly less accurate. The existing gap between 1 and $21.5\,R_{\odot}$ could possibly be filled with MHD models that are able to simulate CME propagation and particle acceleration in the corona \citep[e.g.,][]{li2021,young2021}. An alternative approach may be to take advantage of the `fixed-flare-source' option in SEPMOD, demonstrated by \citet{luhmann2012}. This assumes the presence of a source at the Enlil inner boundary that is able to inject particles through the heliospheric domain. This injection can be synchronised with the onset of a flare and scaled according to its X-ray classification. Such potential improvements will be explored further in the context of real-time space weather forecasts in future work.


\section{Summary and Conclusions} \label{sec:conclusions}

In this work, we have modelled the inner heliospheric context including CMEs and SEPs for a time period including two notable eruptive events in late November and early December 2020. We have employed the WSA--Enlil--SEPMOD modelling chain to simulate the background solar wind and its transients using four different sets of CME input parameters that provide a mini-ensemble framework for predictions of CME arrival and SEP events. CME input parameters were derived from coronagraph observations near Earth and at the STEREO-A spacecraft (located ${\sim}60^{\circ}$ east of the Sun--Earth line), representing observers from the Lagrange L1 and L5 points.

We have found that, despite limitations mostly related to the model's architecture and assumptions, reasonable predictions were obtained for both shock arrivals and SEP event fluxes and durations, at least out to several tens of degrees (${\sim}70^{\circ}$) away from the corresponding eruption site. We noted that CME shock arrival time predictions using real-time parameters (and STEREO beacon data) are only slightly worse than those obtained in hindcasts with science-level data. The modelling setup used is not able to correctly predict SEP event onsets, which may take place before the corresponding CME is estimated to reach the 21.5\,$R_{\odot}$ inner boundary of Enlil. We noted that SEP predictions do not tend to be significantly better for the model run in which the predicted shock arrival times are matched to the observed ones, suggesting that a ``perfect'' CME shock arrival forecast may not be necessary for obtaining reasonable SEP predictions. In fact, the observed variation in the modelled event durations and peak fluxes suggests that an ensemble approach \citep[currently in use in several CME forecasting settings by both NOAA and the UK MET Office; see][]{murray2018} may be beneficial for SEP predictions as well. We conclude that our current real-time forecasting tools can satisfactorily predict CME arrival times as well as SEP peak fluxes and event durations (except the few hours following the event onset) for well-connected observers. The strength of the modelling setup employed here is that WSA--Enlil is already operational for real-time forecasts, whilst SEPMOD is run as a computationally-efficient post-processing procedure on the Enlil simulation output. In fact, SEP predictions from SEPMOD driven in real time are currently displayed on the SEP Scoreboard at NASA's CCMC (\url{https://ccmc.gsfc.nasa.gov/challenges/sep}). Our results indicate that coupling SEPMOD with the existing WSA--Enlil framework in operational settings can provide useful information on the duration and peak fluxes of a gradual SEP event.

We have also shown that a combination of remote-sensing observers placed at the Lagrange L1 and L5 points (i.e., separated by ${\sim}60^{\circ}$) can constrain a CME's propagation direction even over a broad range of heliolongitudes. However, such a configuration is already no longer available, since the STEREO-A spacecraft is currently approaching Earth at a rate of ${\sim}20^{\circ}$ per year until reaching radial alignment in August~2023. By then, the advantages of remote-sensing observations of the Sun from two well-separated viewpoints will be lost at least for a while (we note that SolO is equipped with a coronagraph and a heliospheric imager, but its rapidly-changing heliocentric distance and longitudinal separation with Earth may not be optimal for forecasting purposes). Dedicated missions to the Lagrange L4 and/or L5 points \citep[e.g.,][]{bemporad2021,posner2021,vourlidas2015} may prove beneficial in this regard. This includes the European Space Agency (ESA)'s Vigil \citep{pulkkinen2019} mission, which is currently in development to place a spacecraft equipped with remote-sensing and in-situ instruments at L5. As to in-situ measurements, it is clear that the availability of multi-point observers at different radial distances and heliolongitudes is crucial for analysing and validating modelling results across the entire inner heliosphere. Given the presence of currently-operational spacecraft at six independent locations (i.e., the ones explored in this work), it is important to consider the future opportunities for heliophysics and space weather science via multi-point studies and coordinated observations \citep[e.g.,][]{hadid2021, mostl2022, velli2020}. In conclusion, the potential for significant progress in heliophysics and space weather science will be realised as future studies increasingly utilise the multi-spacecraft capabilities of the Heliophysics System Observatory.


\appendix

\section{WSA--Enlil--SEPMOD Runs} \label{app:runs}

\begin{figure}[p]
\centering
\hspace*{-0.1\linewidth}
\includegraphics[width=1.2\linewidth]{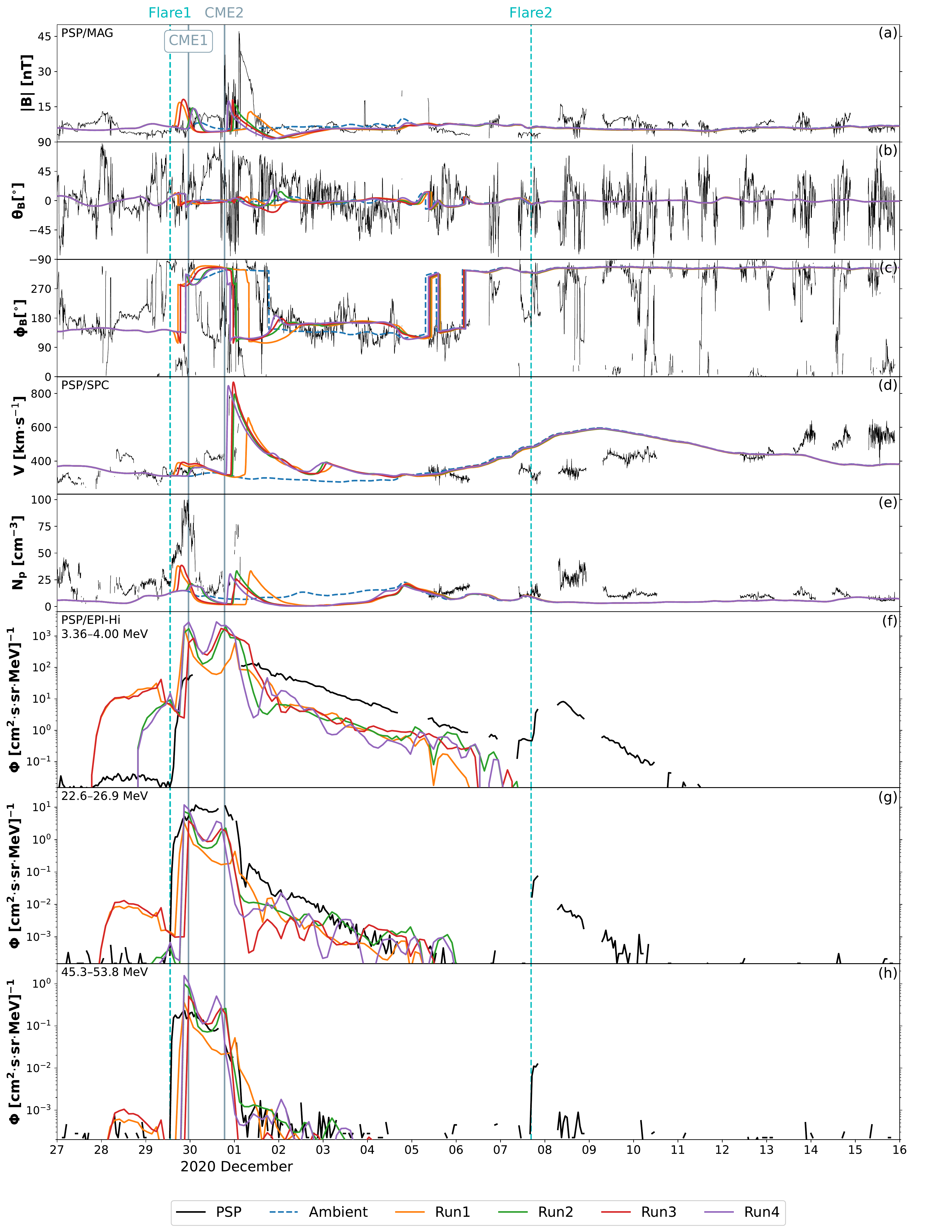}
\caption{WSA--Enlil--SEPMOD results at PSP shown against PSP data. The parameters shown are: (a) magnetic field magnitude, (b) $\theta$ and (c) $\phi$ angles of the magnetic field, (d) solar wind speed, (e) proton density, and (f--h) particle measurements in three different energy channels.}
\label{fig:psp}
\end{figure}

\begin{figure}[p]
\centering
\hspace*{-0.1\linewidth}
\includegraphics[width=1.2\linewidth]{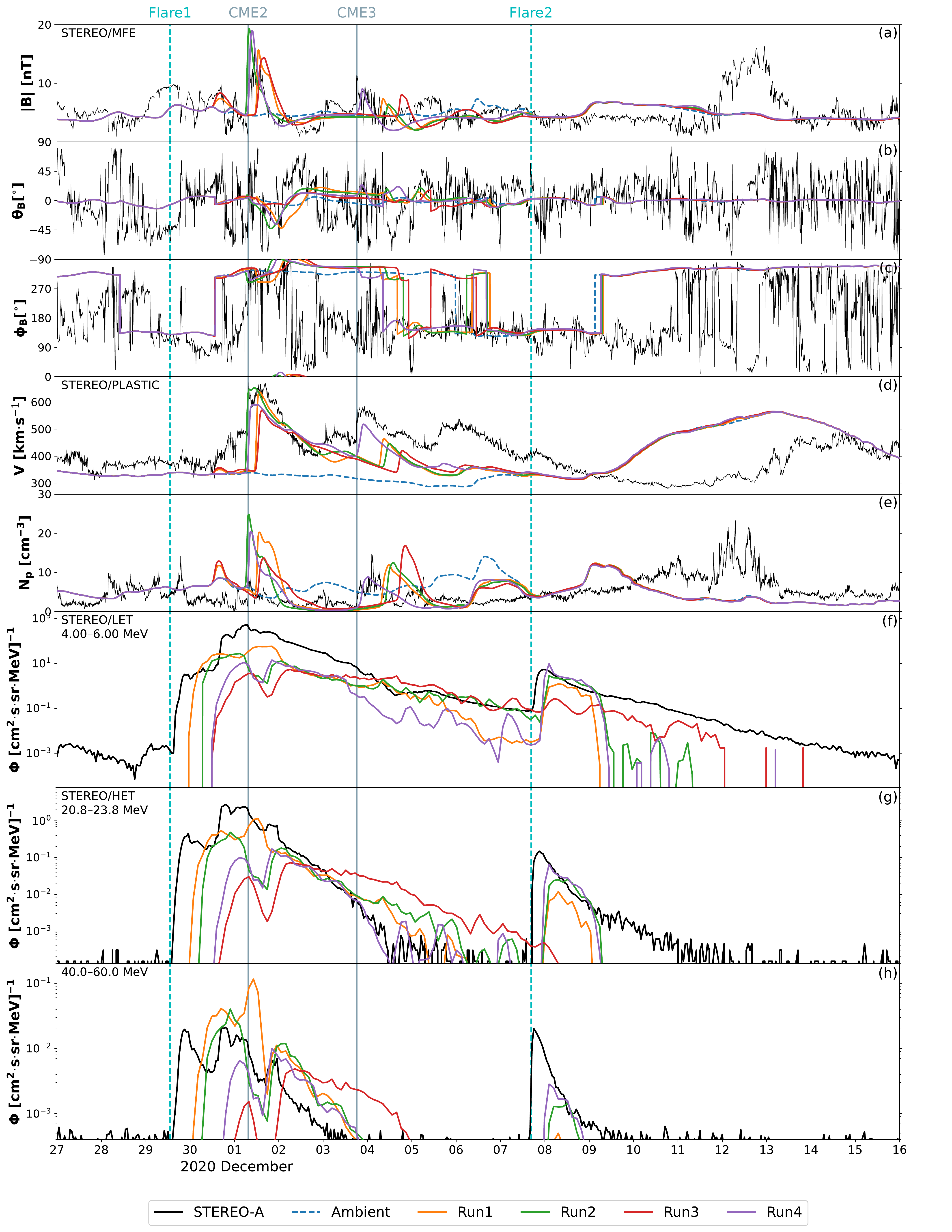}
\caption{WSA--Enlil--SEPMOD results at STEREO-A shown against STEREO-A data. The parameters shown are: (a) magnetic field magnitude, (b) $\theta$ and (c) $\phi$ angles of the magnetic field, (d) solar wind speed, (e) proton density, and (f--h) particle measurements in three different energy channels.}
\label{fig:stereoa}
\end{figure}

\begin{figure}[p]
\centering
\hspace*{-0.1\linewidth}
\includegraphics[width=1.2\linewidth]{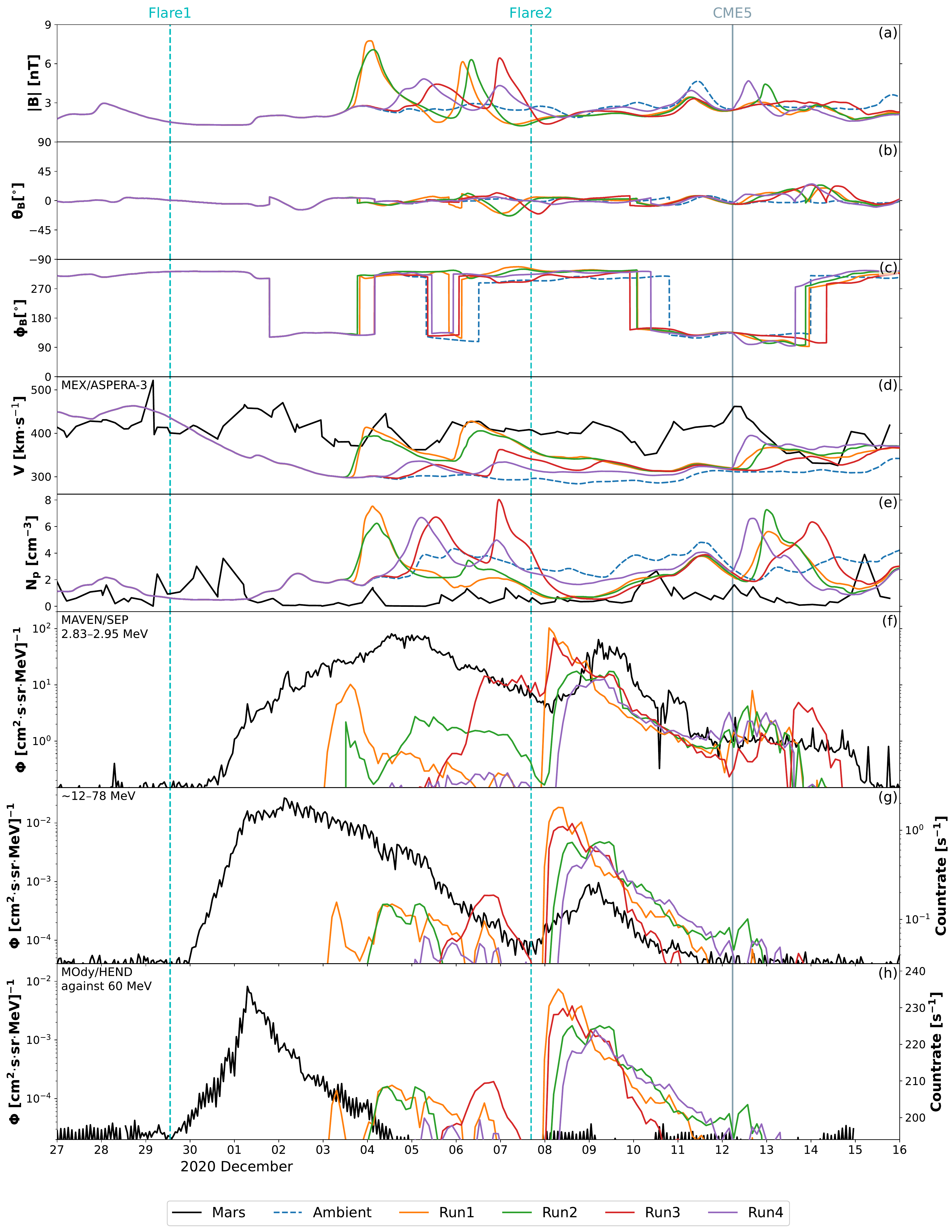}
\caption{WSA--Enlil--SEPMOD results at Mars shown against data from Mars orbit. The parameters shown are: (a) magnetic field magnitude, (b) $\theta$ and (c) $\phi$ angles of the magnetic field, (d) solar wind speed, (e) proton density, and (f--h) particle measurements in three different energy channels.}
\label{fig:mars}
\end{figure}

\begin{figure}[p]
\centering
\hspace*{-0.1\linewidth}
\includegraphics[width=1.2\linewidth]{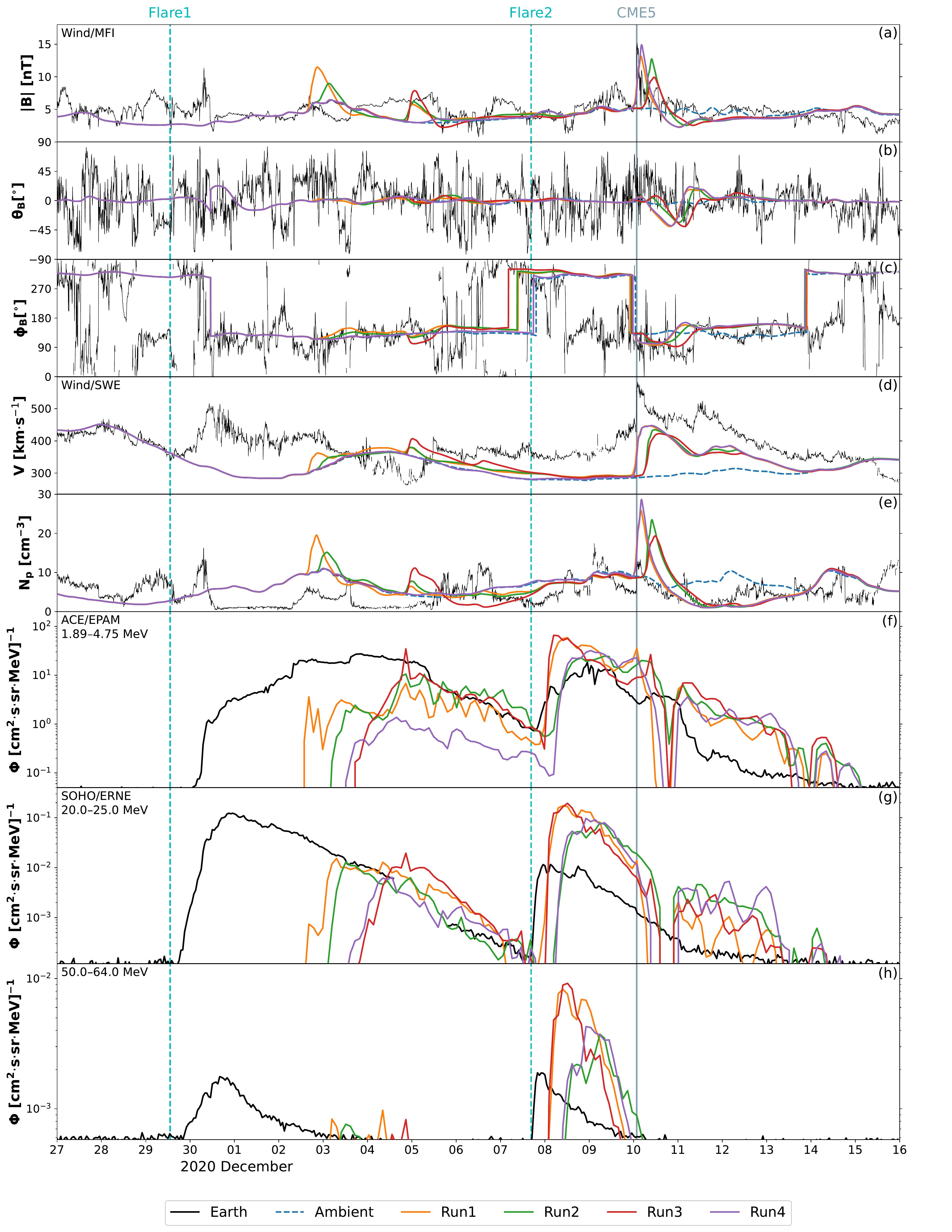}
\caption{WSA--Enlil--SEPMOD results at Earth shown against data taken at Earth's Lagrange L1 point. The parameters shown are: (a) magnetic field magnitude, (b) $\theta$ and (c) $\phi$ angles of the magnetic field, (d) solar wind speed, (e) proton density, and (f--h) particle measurements in three different energy channels.}
\label{fig:earth}
\end{figure}

\begin{figure}[p]
\centering
\hspace*{-0.1\linewidth}
\includegraphics[width=1.2\linewidth]{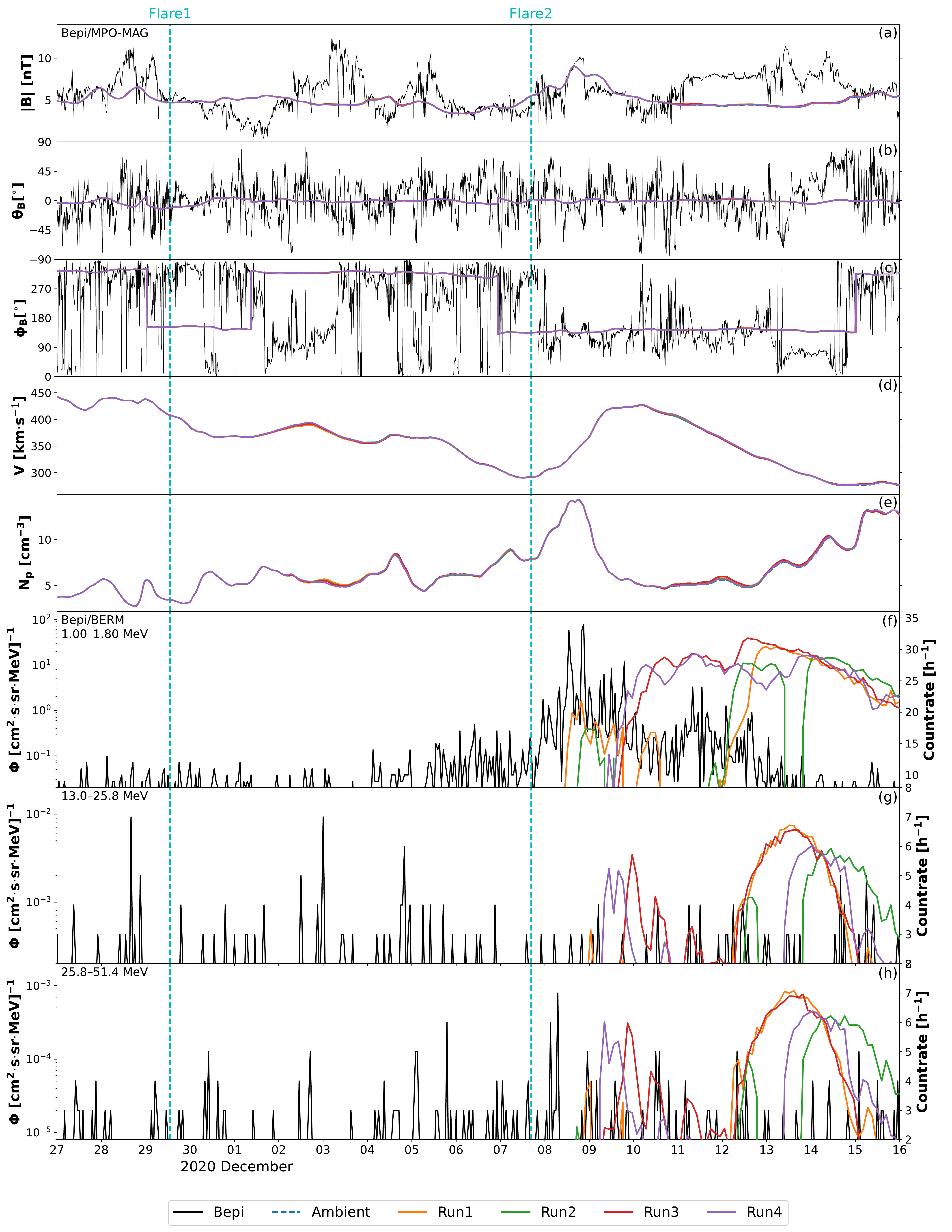}
\caption{WSA--Enlil--SEPMOD results at Bepi shown against Bepi data. The parameters shown are: (a) magnetic field magnitude, (b) $\theta$ and (c) $\phi$ angles of the magnetic field, (d) solar wind speed, (e) proton density, and (f--h) particle measurements in three different energy channels.}
\label{fig:bepi}
\end{figure}

\begin{figure}[p]
\centering
\hspace*{-0.1\linewidth}
\includegraphics[width=1.2\linewidth]{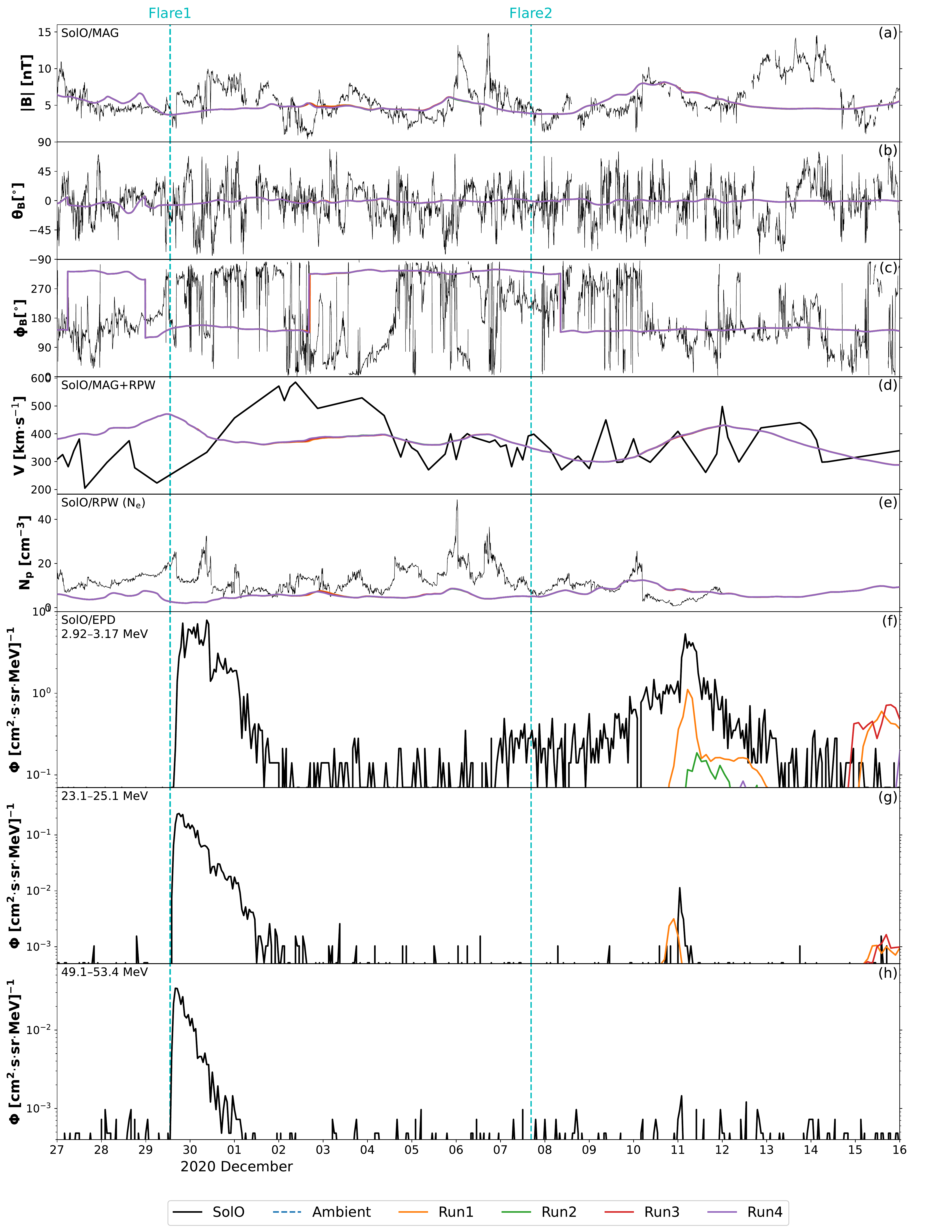}
\caption{WSA--Enlil--SEPMOD results at SolO shown against SolO data. The parameters shown are: (a) magnetic field magnitude, (b) $\theta$ and (c) $\phi$ angles of the magnetic field, (d) solar wind speed, (e) proton density, and (f--h) particle measurements in three different energy channels.}
\label{fig:solo}
\end{figure}

Results of each of the mini-ensemble WSA--Enlil--SEPMOD simulation runs are shown against the corresponding in-situ data in Figures~\ref{fig:psp} to \ref{fig:solo}. Magnetic field and plasma measurements are reported for each observer (where available) and compared directly with Enlil data---including the so-called `ambient' run, which indicates the corresponding run of the solar wind background without CMEs. In each figure, the dashed teal lines mark the peak times of Flare1 and Flare2 (associated with CME2 and CME5, respectively), whilst the solid grey lines mark the arrival in situ of a CME-driven shock (with the driving CME indicated for each case). We remark that speed and density data at Mars (Figure~\ref{fig:mars}) as well as speed data at SolO (Figure~\ref{fig:solo}) are derived rather than directly measured, and are thus expected to be characterised by non-negligible uncertainties. Particle measurements are shown at three illustrative energies (roughly in the ranges 1--6~MeV, 20--25~MeV, and 40--60~MeV) and compared with SEPMOD data that have been interpolated to the centre of each energy channel (indicated to the top left of the corresponding panel). Note that, in each panel, the lower end of the Y-axis has been cut immediately above the corresponding instrument's noise level (i.e., the instrumental background). At PSP (Figure~\ref{fig:psp}), STEREO-A (Figure~\ref{fig:stereoa}), Earth (Figure~\ref{fig:earth}), and SolO (Figure~\ref{fig:solo}), SEPMOD data are directly compared to the measured differential proton fluxes. At Bepi (Figure~\ref{fig:bepi}), SEPMOD fluxes are shown against proton countrates at well-defined energy channels, and the two datasets are arbitrarily normalised with respect to each other. We also remark that Bepi/BERM is placed behind the radiator pointing towards the $-{Y}$ direction, meaning that the instrument is pointing in the anti-sunward direction. This aspect, together with the packed configuration of Bepi during its cruise phase, could imply that the spacecraft itself may screen some particles. Finally, we note that at the moment, only uncalibrated ``raw'' BERM data are available, and a proper calibration is in planning. At Mars (Figure~\ref{fig:mars}), the data in the lowest energy channel are shown in flux units, whilst the remaining two sets of measurements are shown in countrates. The first comes from particles in the Foil--Thick--Open (FTO) detector of MAVEN/SEP, which corresponds to approximately 12--78~MeV protons \citep[more information can be found in][]{larson2015,lee2018}, and is shown against SEPMOD results at 45~MeV (centre of the 12--78~MeV range). The second comes from MOdy/HEND counts that are sensitive to particles of energy as low as 0.2~MeV \citep{sanchezcano2018} up to galactic cosmic rays, and is shown arbitrarily agains SEPMOD results at 60~MeV.

\section{Snapshots from WSA--Enlil} \label{app:snaps}

\begin{figure}[p]
\centering
\begin{adjustwidth}{-1cm}{-1.8cm}
\includegraphics[width=.99\linewidth]{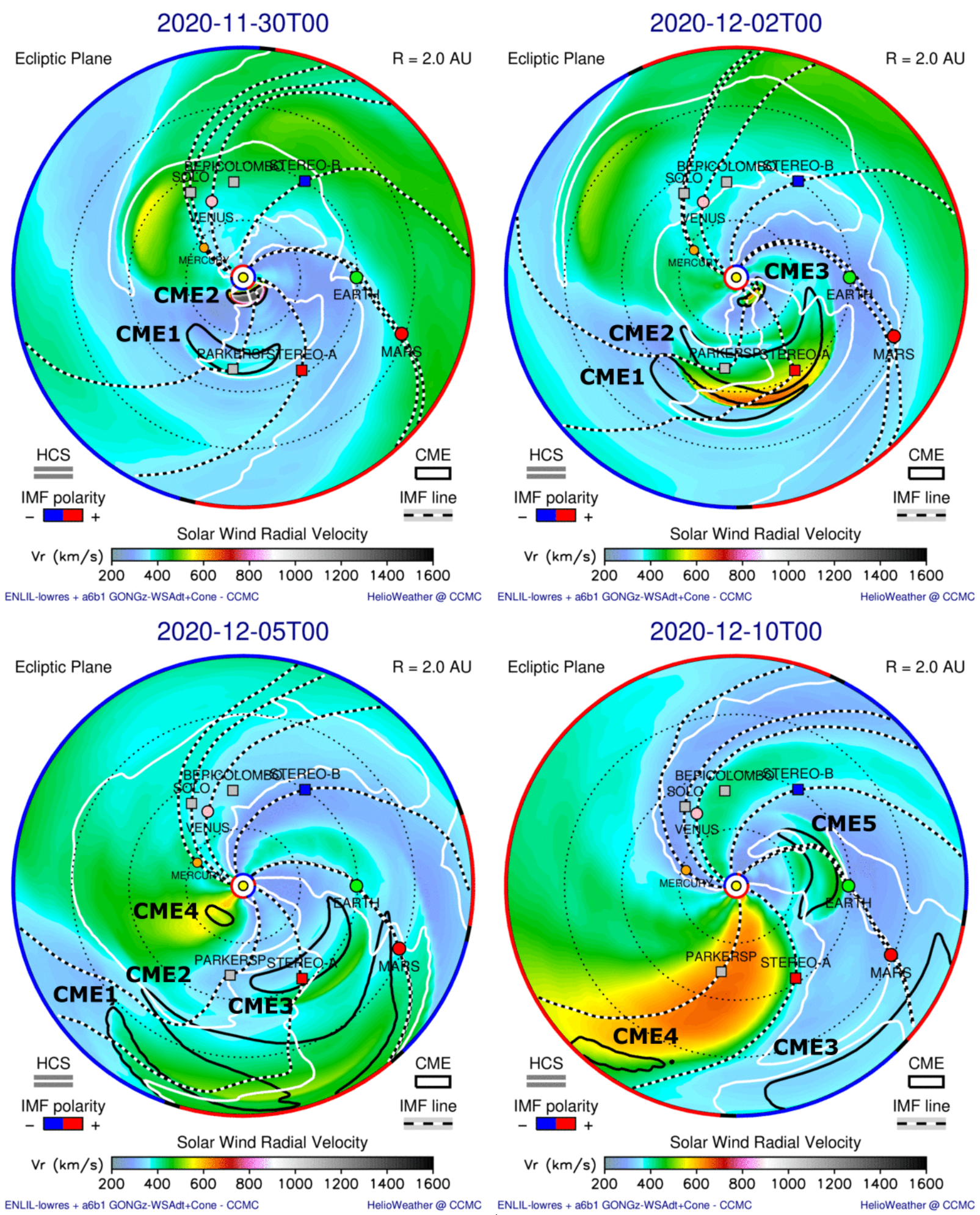}
\caption{Snapshots from the WSA--Enlil simulation for Run1 on the ecliptic plane. The parameter shown is the radial solar wind speed ($V_\mathrm{r}$). From \url{https://ccmc.gsfc.nasa.gov/database_SH/Erika_Palmerio_082921_SH_1.php}.}
\label{fig:run1}
\end{adjustwidth}
\end{figure}

\begin{figure}[p]
\centering
\begin{adjustwidth}{-1cm}{-1.8cm}
\includegraphics[width=.99\linewidth]{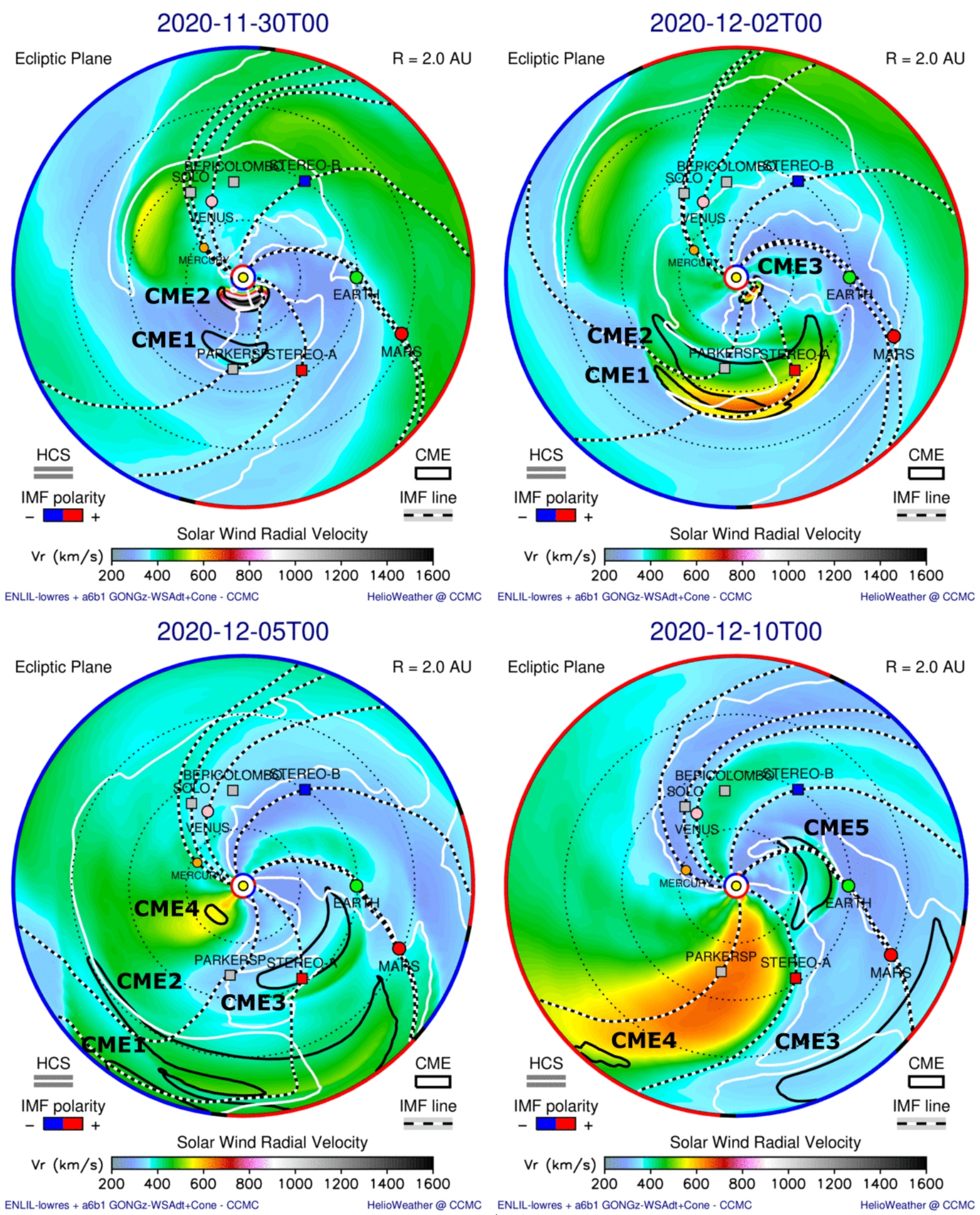}
\caption{Snapshots from the WSA--Enlil simulation for Run2 on the ecliptic plane. The parameter shown is the radial solar wind speed ($V_\mathrm{r}$). From \url{https://ccmc.gsfc.nasa.gov/database_SH/Erika_Palmerio_082921_SH_2.php}.}
\label{fig:run2}
\end{adjustwidth}
\end{figure}

\begin{figure}[p]
\centering
\begin{adjustwidth}{-1cm}{-1.8cm}
\includegraphics[width=.99\linewidth]{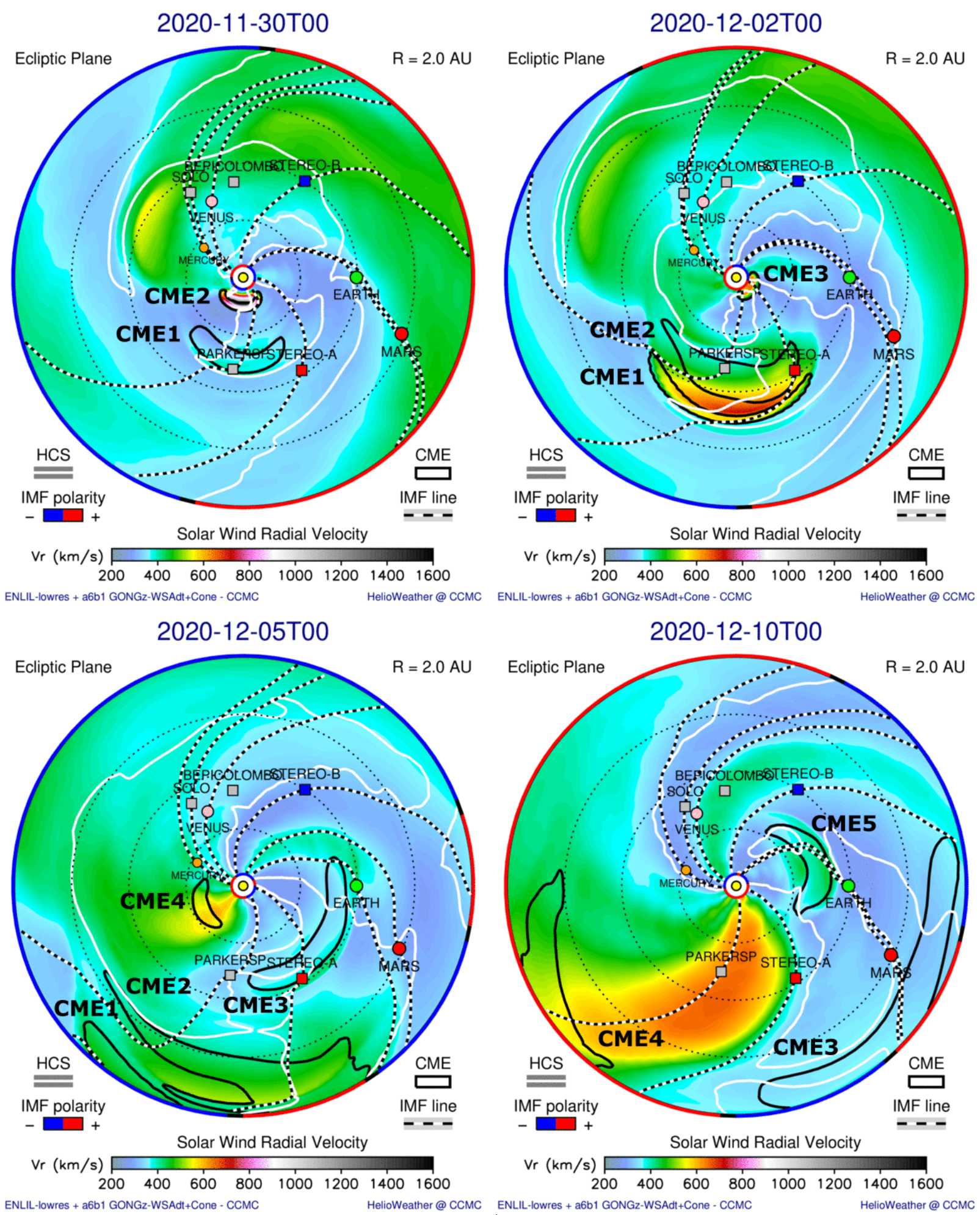}
\caption{Snapshots from the WSA--Enlil simulation for Run3 on the ecliptic plane. The parameter shown is the radial solar wind speed ($V_\mathrm{r}$). From \url{https://ccmc.gsfc.nasa.gov/database_SH/Erika_Palmerio_082921_SH_3.php}.}
\label{fig:run3}
\end{adjustwidth}
\end{figure}

\begin{figure}[p]
\centering
\begin{adjustwidth}{-1cm}{-1.8cm}
\includegraphics[width=.99\linewidth]{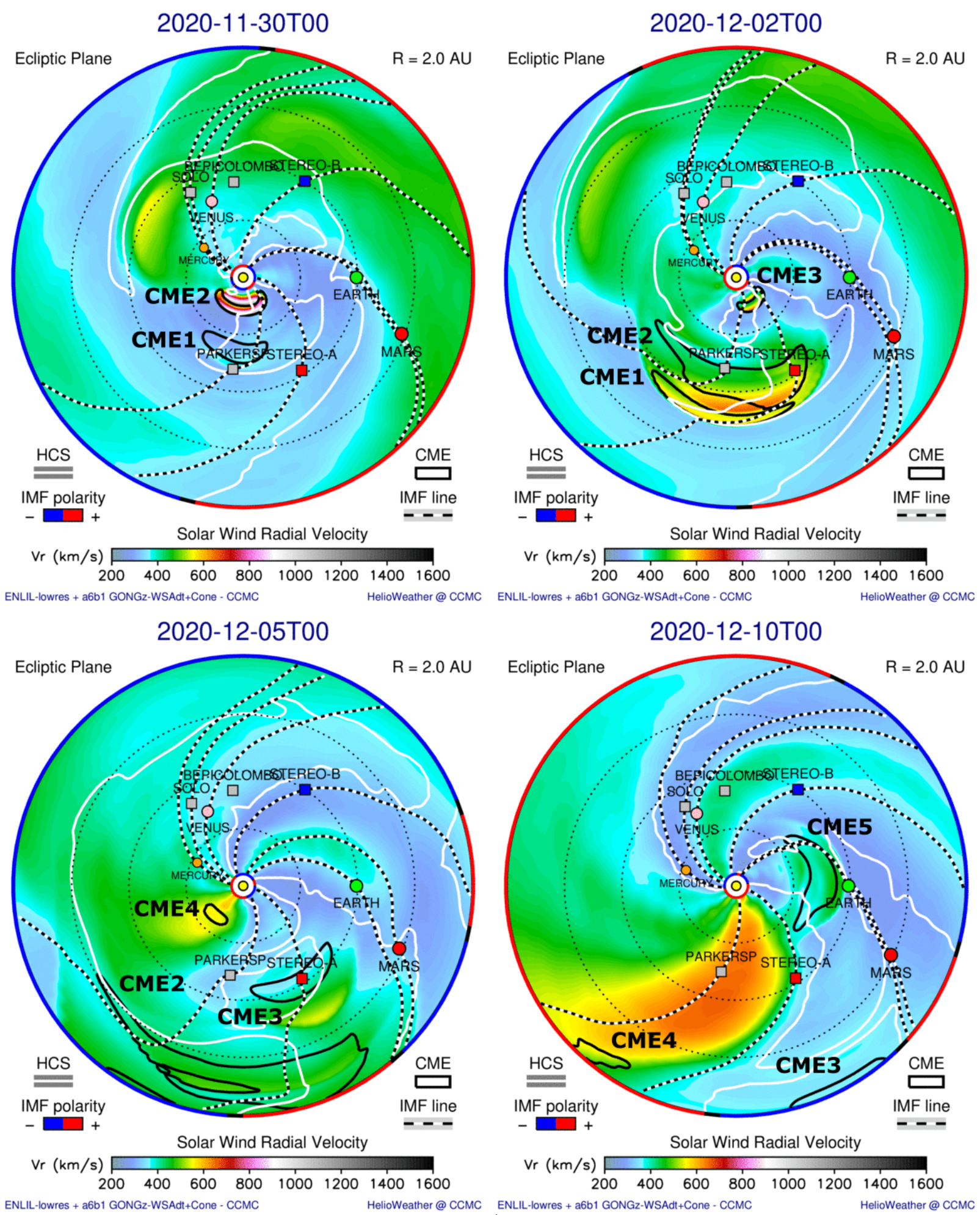}
\caption{Snapshots from the WSA--Enlil simulation for Run4 on the ecliptic plane. The parameter shown is the radial solar wind speed ($V_\mathrm{r}$). From \url{https://ccmc.gsfc.nasa.gov/database_SH/Erika_Palmerio_082921_SH_4.php}.}
\label{fig:run4}
\end{adjustwidth}
\end{figure}

Snapshots from the WSA--Enlil simulation runs are shown in Figures~\ref{fig:run1} to \ref{fig:run4}. Each figure shows the modelled radial solar wind speed on the ecliptic plane at four different times. The five CMEs analysed in this study are labelled in each panel.

\section*{Data Availability Statement}

Simulation results have been provided by the Community Coordinated Modeling Center at Goddard Space Flight Center through their public Runs on Request system (\url{http://ccmc.gsfc.nasa.gov}).

DONKI runs used in this study:
\begin{itemize}
\item CME1: \url{https://kauai.ccmc.gsfc.nasa.gov/DONKI/view/CMEAnalysis/16146/2}
\item CME2: \url{https://kauai.ccmc.gsfc.nasa.gov/DONKI/view/CMEAnalysis/16172/1}
\item CME3: \url{https://kauai.ccmc.gsfc.nasa.gov/DONKI/view/CMEAnalysis/16186/3}
\item CME4: \url{https://kauai.ccmc.gsfc.nasa.gov/DONKI/view/CMEAnalysis/16204/2}
\item CME5: \url{https://kauai.ccmc.gsfc.nasa.gov/DONKI/view/CMEAnalysis/16219/2}
\end{itemize}

WSA--Enlil--SEPMOD runs performed in this study:
\begin{itemize}
\item Run1:
\begin{itemize}
\item Enlil: \url{https://ccmc.gsfc.nasa.gov/results/viewrun.php?domain=SH&runnumber=Erika_Palmerio_082921_SH_1}
\item SEPMOD (GOES-like): \url{https://ccmc.gsfc.nasa.gov/results/viewrun.php?domain=SH&runnumber=Erika_Palmerio_090121_SH_1}
\item SEPMOD (IMP8-like): \url{https://ccmc.gsfc.nasa.gov/results/viewrun.php?domain=SH&runnumber=Erika_Palmerio_090121_SH_2}
\end{itemize}
\item Run2:
\begin{itemize}
\item Enlil: \url{https://ccmc.gsfc.nasa.gov/results/viewrun.php?domain=SH&runnumber=Erika_Palmerio_082921_SH_2}
\item SEPMOD (GOES-like): \url{https://ccmc.gsfc.nasa.gov/results/viewrun.php?domain=SH&runnumber=Erika_Palmerio_090121_SH_3}
\item SEPMOD (IMP8-like): \url{https://ccmc.gsfc.nasa.gov/results/viewrun.php?domain=SH&runnumber=Erika_Palmerio_090121_SH_4}
\end{itemize}
\item Run3:
\begin{itemize}
\item Enlil: \url{https://ccmc.gsfc.nasa.gov/results/viewrun.php?domain=SH&runnumber=Erika_Palmerio_082921_SH_3}
\item SEPMOD (GOES-like): \url{https://ccmc.gsfc.nasa.gov/results/viewrun.php?domain=SH&runnumber=Erika_Palmerio_090121_SH_5}
\item SEPMOD (IMP8-like): \url{https://ccmc.gsfc.nasa.gov/results/viewrun.php?domain=SH&runnumber=Erika_Palmerio_090121_SH_6}
\end{itemize}
\item Run4:
\begin{itemize}
\item Enlil: \url{https://ccmc.gsfc.nasa.gov/results/viewrun.php?domain=SH&runnumber=Erika_Palmerio_082921_SH_4}
\item SEPMOD (GOES-like): \url{https://ccmc.gsfc.nasa.gov/results/viewrun.php?domain=SH&runnumber=Erika_Palmerio_090121_SH_7}
\item SEPMOD (IMP8-like): \url{https://ccmc.gsfc.nasa.gov/results/viewrun.php?domain=SH&runnumber=Erika_Palmerio_090121_SH_8}
\end{itemize}
\end{itemize}

Remote-sensing data from SDO, SOHO, and STEREO are openly available at the Virtual Solar Observatory (VSO; \url{https://sdac.virtualsolar.org/}). These data were processed and analysed trough SunPy \citep{sunpy2020}, IDL SolarSoft \citep{freeland1998}, and the ESA JHelioviewer software \citep{muller2017}. 

PSP, STEREO, Wind, and ACE in-situ data are publicly available at NASA's Coordinated Data Analysis Web (CDAWeb) database (\url{https://cdaweb.sci.gsfc.nasa.gov/index.html/}). MAVEN data can be accessed at the Planetary Plasma Interactions (PPI) Node of NASA's Planetary Data System (PDS; \url{https://pds-ppi.igpp.ucla.edu}). MEX data are stored at at ESA's Planetary Science Archive (PSA; \url{https://archives.esac.esa.int/psa}). MOdy data are available at the Geosciences Node of the PDS (\url{https://pds-geosciences.wustl.edu/}). SOHO particle data can be downloaded from the SOHO/ERNE webpage (\url{https://srl.utu.fi/projects/erne/}). The Bepi data used in this study are stored at \url{https://doi.org/10.25392/leicester.data.16941040}, and the full mission data set will be released in the future at ESA's PSA. SolO data are openly available at ESA's Solar Orbiter Archive (\url{http://soar.esac.esa.int/soar/}).


\acknowledgments
E.~Palmerio's research was supported by the NASA Living With a Star (LWS) Jack Eddy Postdoctoral Fellowship Program, administered by UCAR's Cooperative Programs for the Advancement of Earth System Science (CPAESS) under award no. NNX16AK22G.
C.~O.~Lee acknowledges support from NASA MAVEN Project subcontract via Grant no. NNH10CC04C, managed by the University of Colorado. C.~O.~Lee and J.~G.~Luhmann acknowledge support from the STEREO Project via the STEREO/IMPACT grant to UC Berkeley.
D.~Lario and I.~G.~Richardson acknowledge support from NASA LWS programs NNH17ZDA001N-LWS and NNH19ZDA001N-LWS, as well as the Goddard Space Flight Center Heliophysics Innovation Fund (HIF) program.
B.~S{\'a}nchez-Cano acknowledges support through UK-STFC Ernest Rutherford Fellowship ST/V004115/1 and STFC grants ST/S000429/1 and ST/V000209/1.
K.~Steinvall and Y.~V.~Khotyaintsev acknowledge support from the Swedish Research Council, grant 2016-05507, and the Swedish National Space Agency (SNSA), grant 20/136.
C.~M{\"o}stl and A.~J.~Weiss thank the Austrian Science Fund (FWF): P31521-N27, P31659-N27.
The WSA model was developed by C.~N.~Arge (currently at NASA/GSFC), the Enlil model was developed by D.~Odstrcil (currently at GMU), and the SEPMOD model was developed by J.~Luhmann (currently at UCB).
Finally, we thank the Mission Teams of all the spacecraft and instruments employed in this study. This work was supported by NASA’s Parker Solar Probe Mission, contract no. NNN06AA01C. Parker Solar Probe was designed, built, and is now operated by the Johns Hopkins Applied Physics Laboratory as part of NASA’s LWS program.

\bibliography{bibliography}

\end{document}